\documentclass{aastex631}
\usepackage{amssymb}
\usepackage{amsmath}
\usepackage{savesym} 
\savesymbol{tablenum}
\usepackage{siunitx}
\restoresymbol{SIX}{tablenum}
\usepackage[online]{threeparttablex}
\usepackage{longtable,booktabs}

\definecolor{DarkGreen}{rgb}{0.0,0.4,0.0}  

\def\degree{$^{\circ}$}

\begin{document}
\title{What Causes the Asymmetry of Conjugate Hard X-Ray Footpoints in Solar Flares?}

\author[0009-0003-9088-4697]{Mirabbos Mirkamalov}
\affiliation{CAS Key Laboratory of Geospace Environment, Department of Geophysics and Planetary Sciences, University of Science and Technology of China, Hefei, Anhui 230026, People's Republic of China}
\affiliation{Department of Nuclear Physics and Astronomy, Institute of Nuclear Technologies, Samarkand State University, Samarkand 140104, Republic of Uzbekistan}

\author[0000-0003-4618-4979]{Rui Liu}
\affiliation{CAS Key Laboratory of Geospace Environment, Department of Geophysics and Planetary Sciences, University of Science and Technology of China, Hefei, Anhui 230026, People's Republic of China}

\author[0000-0002-2559-1295]{Runbin Luo}
\affiliation{CAS Key Laboratory of Geospace Environment, Department of Geophysics and Planetary Sciences, University of Science and Technology of China, Hefei, Anhui 230026, People's Republic of China}

\author[0000-0001-5313-1125]{Arun Kumar Awasthi}
\affiliation{Department of Physics and Astrophysics, University of Delhi, Delhi, 110007, India}
\affiliation{Space Research Centre, Polish Academy of Sciences, Bartycka 18A, 00-716 Warsaw, Poland}

\correspondingauthor{Rui Liu}
\email{rliu@ustc.edu.cn}

\begin{abstract}

Hard X-ray (HXR) emission in solar flares critically diagnoses nonthermal electron acceleration, transport, and precipitation. Observations commonly show asymmetric HXR photon fluxes between paired footpoints, whose physical origin remains debated, as the conventional magnetic mirroring mechanism often fails to explain the observed asymmetry. Here we performed rigorous statistical tests on the association between photospheric magnetic parameters within the HXR footpoint regions and the asymmetry of HXR production. We analyzed 103 time intervals taken from around the peaks of HXR bursts in 67 M- and X-class flares with a clear double-ribbon morphology observed by both the Ramaty High Energy Solar Spectroscopic Imager and the Solar Dynamics Observatory. We found that conjugate HXR footpoint sources are asymmetric in photon fluxes, maximum intensities, and sizes, but rather symmetric in mean intensities. The photon flux ratio of the stronger over weaker footpoint source shows a strong linear correlation with the size ratio and a nonlinear correlation with the maximum intensity ratio. The asymmetry of magnetic field strength and flux at the conjugate footpoints shows a positive correlation with the HXR asymmetry, contrary to what magnetic mirroring effects predict. Importantly, the asymmetry of unsigned photospheric vertical electric current (PVEC) exhibits a strong positive correlation with the HXR footpoint asymmetry. PVEC at HXR footpoints most likely maps the footprints of coronal current layers where flaring reconnections occur. This tight linkage suggests that the electric-current-associated physical processes, including reconnection-induced electric fields and current-driven micro-turbulence, are at work to modulate the production and precipitation of nonthermal electrons. 


\end{abstract}

\section{Introduction} \label{sec:intro}

In the standard flare model, explosive energy release during solar flares is dictated by magnetic reconnection occurring in the corona \citep{1964NASSP..50..451C,1966Natur.211..695S,1974SoPh...34..323H,1976SoPh...50...85K}, resulting in energetic phenomena such as plasma heating, mass ejection, and particle acceleration. Accelerated in the tenuous corona to energies as high as tens to hundreds of keV, beams of nonthermal electrons propagate along newly reconnected magnetic field lines downward to bombard the dense atmospheric layers, where they get collisionally stopped and thermalized. Part of their energy is converted via the thick-target bremsstrahlung process to HXR emission in the transition region and chromosphere. The HXR radiation contains the most ``direct'' information about the acceleration and transportation of nonthermal electrons \citep{1971SoPh...18..489B,2003ApJ...595L..97H,2005A&A...435..743S}, and has been detected in increasingly rich details by a succession of spacecraft missions such as the Solar Maximum Mission in the 1980s \citep{1984AdSpR...4b.153T}, Yohkoh in the 1990s, Reuven Ramaty High Energy Solar Spectroscopic Imager \cite[RHESSI;][]{2002SoPh..210....3L} from 2002 to 2018, and more recently the Solar Orbiter and the Advanced Space-based Solar Observatory.

The HXR images of flares typically show a double source associated with two flare ribbons located on either side of the polarity inversion line (PIL). The double source are generally interpreted as the precipitation sites of nonthermal electrons at magnetically conjugate footpoints of a flare loop \cite[]{Holman2011}. The conjugate HXR footpoints exhibit strong correlation in their light curves \citep{Jin&Ding2007} and resemblance in spectral index \citep{Saint-Hilaire2008}, indicating that they are produced by the same population of accelerated electrons; paradoxically, they also exhibit imbalanced HXR fluxes \citep{Sakao1994PhDT,Goff2004,Yang2012}. Theoretically, the asymmetry is regulated by the magnetic field configuration and plasma conditions of the flaring loop and involves multiple physical processes, from the acceleration and injection of nonthermal electrons in the corona, to the streaming of the electrons along the two legs of the flaring loop, which is impacted by collisions, ohmic heating, as well as the self-induced electric field, and eventually to the deposition of the electron energy at the conjugate footpoints \cite[]{Aschwanden2002}. A straightforward interpretation invokes the conservation of the magnetic moment ($\mu\propto \sin^2\alpha/B$) of a free-streaming electron with the pitch angle $\alpha$. As the magnetic field typically converges approaching the photosphere, this electron is reflected back at the mirroring point where the field strength reaches $B_0/\sin^2\alpha$, with $B_0$ being the field strength at the injection point. The rate of electron precipitation and the resultant thick-target HXR flux are then expected to be higher at the footpoint with the weaker magnetic field strength.

The association in asymmetry between the HXR fluxes and the photospheric magnetic field strengths at the conjugate footpoints was first studied by \cite{Sakao1994PhDT} and investigated extensively by follow-up studies \cite[e.g.,][]{Kundu1995, Sakao1996, Aschwanden1999,Li1997,Asai2002,Goff2004,Siarkowski&Falewicz2004,Falewicz&Siarkowski2007,Yang2012,Daou&Alexander2016}. Two types of footpoint asymmetry with respect to the mean field strength are often named after the work of \cite{Sakao1994PhDT}: S-type (Sakao type) events have a brighter HXR FP in the region of weaker photospheric magnetic field; the reverse is true for N-type (non-Sakao type). Although S-type events are consistent with the magnetic mirroring effect, N-type events are discovered in a significant fraction of flares. \cite{Goff2004} reported N-type asymmetry in 11 of 32 flares. \cite{Yang2012} studied 172 time intervals in 22 flares and reported S-type asymmetry for 75\% of the cases. Obviously, other effects in addition to magnetic mirroring, such as asymmetric injection site of accelerated electrons \cite[e.g.,][]{Goff2004,Falewicz&Siarkowski2007} or the different plasma densities in a flaring loop \cite[e.g.,][]{Falewicz&Siarkowski2007,Liu2009}, must have also played significant roles. Notably, the asymmetric behavior of HXR footpoints is also found to exhibit energy-dependence and vary with time \citep{Alexander&Metcalf2002, Siarkowski&Falewicz2004, McClements&Alexander2005, Falewicz2006, Liu2009, Yang2012}. \cite{Daou&Alexander2016} reported that the HXR footpoint asymmetry saturates when the magnetic field ratio of conjugate footpoints exceeds about 4. These phenomena prompted new questions about the role of magnetic fields, plasma conditions, as well as particle acceleration and injection processes in producing an imbalance in HXR footpoint emission.

In contrast to the relatively compact HXR footpoints, the footpoint regions of the flaring loops in SXR/EUV often take the shape of two elongated bright ribbons in UV in the transition region or chromosphere \citep{Fletcher2011}, which is considered a direct observational feature of energy deposition in the lower layers of the solar atmosphere. Sometimes the two opposite far ends of the conjugated flare ribbons exhibit a hooked morphology, which outlines the footpoints of the eruptive magnetic flux rope \cite[e.g.,][]{Wang2017,Gou2023}. These observations are consistent with the 3D extension of the 2D standard flare model \citep{Janvier2014,Janvier2017}. In this model, the magnetic flux rope is wrapped around by a quasi-separatrix layer (QSL), whose footprints correspond to the hooked segments of flare ribbons, and the vertical current sheet underneath the flux rope in the classical 2D model is embedded within two intersecting QSLs, known as a hyperbolic flux tube (HFT), whose footprints correspond to the straight segments of flare ribbons. Analogous to but different from separatrix surfaces, QSLs and HFTs are continuous geometric structures associated with rapid changes in magnetic connectivity, where current layers tend to form. Thus, the spatial correlation between the flare ribbons and the channels of photospheric vertical electric currents (PVECs) is taken as a strong evidence for the 3D standard model. \cite{He2020} found that flares with J-shaped ribbons are more likely to be associated with CMEs than those with non-J-shaped ribbons. 

Motivated by the 2011 February 15 flare \citep{Janvier2014,Musset2015}, in which the brighter EUV/UV flare ribbon as well as the brighter HXR footpoint is associated with stronger upward electric currents, \cite{Haerendel2017} suggests that the observation provides evidence for electron acceleration by field-aligned potential drops generated by anomalous resistivity of highly filamentary currents whose magnitude is on the order of $10^4$ A$\,$m$^{-2}$. Although such extremely dense currents are undetectable with the resolution of modern instruments, the strong photospheric currents indicate the presence of magnetic shear, which may supply the free energy needed for the acceleration process. However, earlier observational studies on the role of PVECs in particle acceleration seem to be inconclusive and sometimes at odds with \cite{Haerendel2017}. Early ground-based optical observations found that in many flares a large percentage of H$\alpha$ emission sources are in proximity (within $6''
$) to local PVEC maxima or overlap with them \cite[see][for a review]{Zimovets&Sharykin2022}. Using H$\alpha$ Stark-wing emission as a proxy, \cite{Canfield1993b} and \cite{Canfield1993c} found that energetic electron precipitation tends to occur near regions of strong vertical currents. Using HXRs as a direct diagnostic of particle precipitation, \cite{Li1997} concluded that electron precipitation tends to avoid sites of strong vertical currents but be preferentially adjacent to these current channels; they also found that at magnetically conjugate footpoints, stronger HXR emission is associated with smaller vertical current density and weaker magnetic field. 

The inconclusiveness on the relation between HXR footpoints and PVECs may arise from the small number of flares investigated, due mainly to the limited availability of vector magnetograms before the launch of the Solar Dynamics Observatory \cite[SDO;][]{Pesnell2012} in 2010. The Helioseismic and Magnetic Imager \cite[HMI;][]{2012SoPh..275..207S,2012SoPh..275..229S} onboard SDO has made available the regular measurement of the photospheric vector magnetic field and therefore vertical currents in high precision and spatiotemporal resolution. However, much of the attention has been focused on the relation between flare ribbons and PVECs in individual events \cite[e.g.,][]{Janvier2014, Sharykin&Kosovichev2014,Sharykin_2015,Musset2015,Sharykin2020}. \cite{Zimovets2020apj} presented the first statistical study on the relation between flare HXR sources and PVECs, with a sample of 48 flares during 2010--2015. They confirmed that the 50--100 keV HXR sources tend to be located at the periphery of local PVEC maxima, but found no significant correlation between the HXR source intensity and PVEC density or total PVEC under the sources. 

Within the framework of the standard flare model, the HXR footpoint asymmetry is a reflection of different magnetic and plasma conditions along the two legs of the newly reconnected flaring loop, thereby offering insight into the physical processes that regulate particle acceleration and energy deposition in solar flares. This motivates a systematic investigation on how the asymmetry of HXR emission relates to photospheric magnetic fields and electric currents, which is the focus of the present work. Through a comprehensive survey covering the entire observational overlap period between SDO and RHESSI during 2010--2018, we collect a large sample of flares suitable for this study. We define the HXR footpoint asymmetry as the imbalance between the two conjugate footpoints not only in photon flux as in the conventional approach, but also in maximum \emph{intensity} (used interchangeably with \emph{brightness}) and source area. Motivated by \cite{Haerendel2017}, we investigate both the sign and magnitude of PVECs. The rest of the paper is organized as follows. In \S\ref{sec:methods} we elaborate on the methodology of the study including the selection of events, the identification of HXR footpoints, and the mapping of photospheric events. We then specify the statistical results in \S\ref{sec:statistics}, discuss their implications in \S\ref{sec:discussion}, and finally give concluding remarks in \S\ref{sec:conclusion}.

\section{Methods} \label{sec:methods}

In this section, we describe the instruments used in this study (\S\ref{subsec:instr}), how we select our events -- the double-ribbon flares (\S\ref{subsec:event_select}), how we reconstruct and identify the magnetically conjugate HXR footpoint sources (\S\ref{subsec:FP_indentify}), and prepare time-averaged PVEC maps (\S\ref{subsec:PVEC}). 

\subsection{Instruments} \label{subsec:instr}

Launched on 11 February 2010 and currently in operation, SDO provides continuous full-disk observations of the Sun and comprises three instruments. Among them, the HMI \citep{2012SoPh..275..207S, 2012SoPh..275..229S} observes Doppler velocity, magnetic flux, continuum intensity, and measures vector magnetic fields in the \ion{Fe}{1} absorption line at 6173~{\AA} with a pixel scale of $0''.5$, and time cadence of 45 seconds for line-of-sight magnetograms. For vector magnetograms, the cadence is 12 minutes due to the pre-inversion average of the four Stokes profiles to improve the signal-to-noise ratio. We used the magnetograms before the flare impulsive phase to avoid the complications of precipitating high-energy electrons impacting on Stokes profiles \cite[]{Qiu&Gary2003}. In this way, we ignore any transient or permanent changes of magnetic field at the HXR footpoints during energetic flares \cite[]{Burtseva2015} and leave this issue for future investigations. The Atmospheric Imaging Assembly \citep[AIA;][]{2012SoPh..275...17L,2012SoPh..275...41B} on board SDO has four identical telescopes providing high temporal (cadence of 12 s for EUV and 24 s for UV filters) and spatial resolution ($1''.5$ with a $0''.6$ pixel size), with nine passbands in the EUV/UV wavelengths and one in visible wavelengths.  

Launched on 5 February 2002 and ceased science operations on 11 April 2018, RHESSI offered an unprecedented combination of spectral resolution (1 keV in the 3-100 keV range), spatial resolution ($2''.3$), temporal evolution ($\approx2$ s), and sensitivity for observations of flare HXR emission for almost two decades. In our work we use RHESSI data to reconstruct flare HXR footpoints at 25--50, 50--100, and 100--300 keV energy ranges. The conjunction of HMI polarimetric measurements in active regions (ARs) with HXR imaging capabilities of RHESSI for flares provides a possibility to study systematically the asymmetry of conjugate footpoints and how the sites of electron precipitation are related to photospheric magnetic fields and electric currents. 

\subsection{Selection of Events} \label{subsec:event_select}

\begin{figure}[ht]
  \centering
  \includegraphics[scale=0.48]{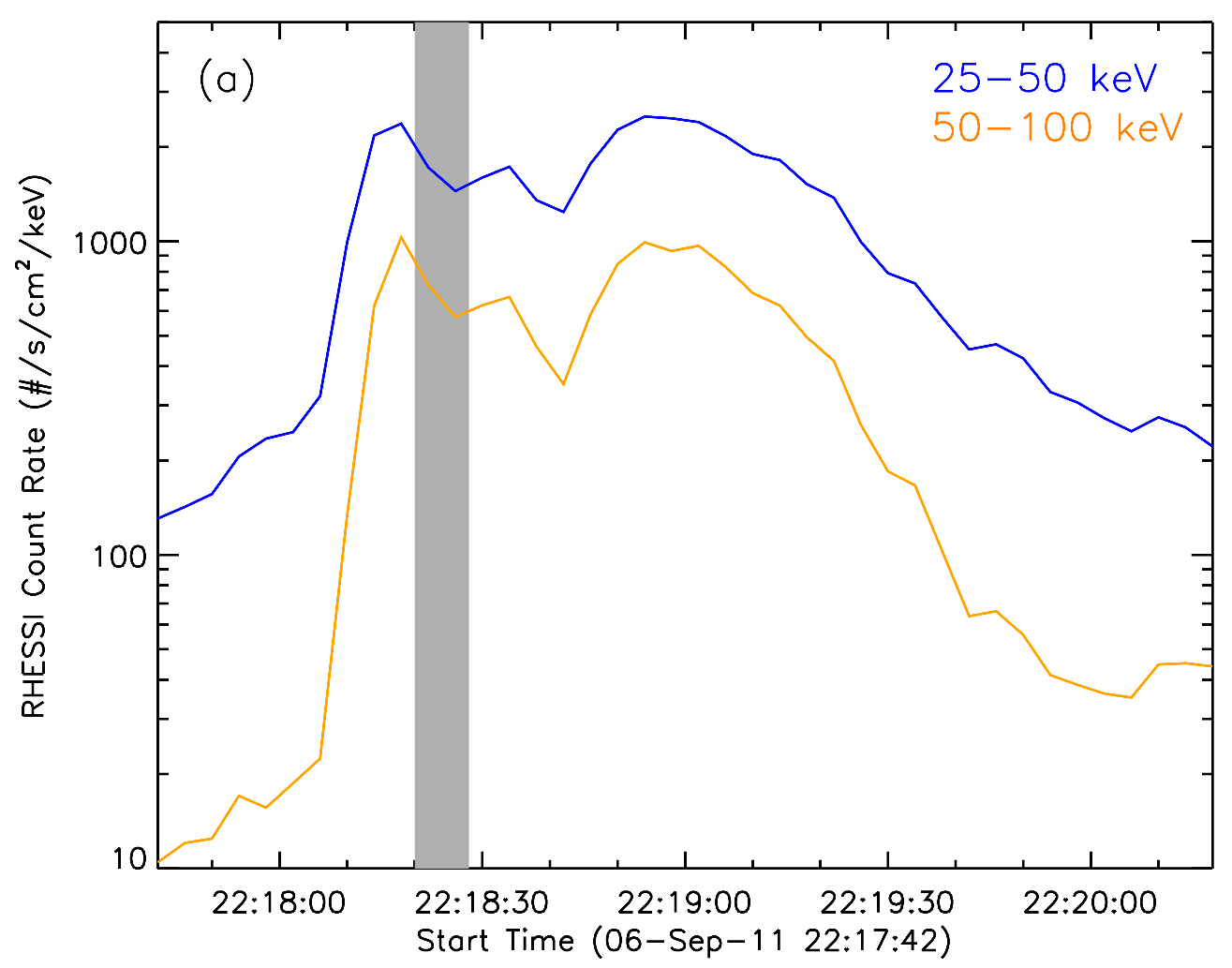} 
  \includegraphics[scale=0.59]{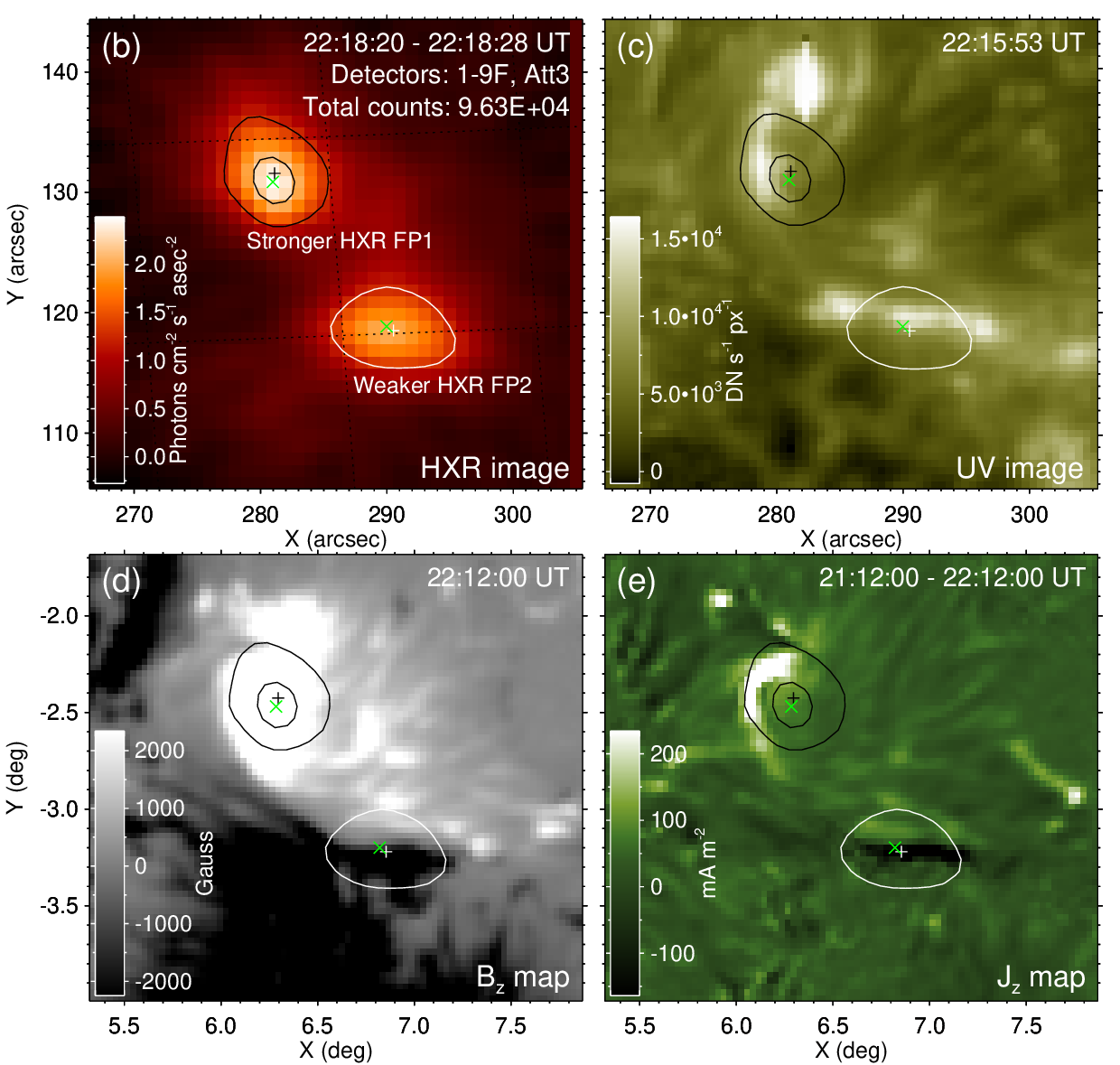}
  \caption{An example GOES X2.1-class double-ribbon flare observed on 2011 September 6 in NOAA AR 11283. (a) Intensity profiles in terms of HXR corrected count rate in two energy bands; the grey shaded area indicates the selected time interval (No.6 in Table~\ref{tab:events}) used to reconstruct the HXR image at 25-50 keV in CCD coordinates in (b). (b--e) The contours of 25-50 keV images at the 50\% and 85\% levels, overlaid on the co-temporal reconstructed X-ray image (b), 1600~{\AA} image (c), as well as the map of $B_z$ (d) and $J_z$ (e) in CEA coordinates. Black (white) contour represents the stronger (weaker) footpoint. The centroid and brightest pixel of each HXR source are marked by `+' and `×' symbols, respectively.
  Noticeably, the HXR footpoints are located in regions of opposite polarities and of strong $J_z$ with opposite signs.    
}
  \label{fig:example}
\end{figure}

In this study, we selected solar flares of GOES-class M1.0 and above, so that the HXR footpoints can be well observed by RHESSI. Among these flares, we chose those with a double-ribbon, i.e., the two ribbons are magnetically conjugated. The identification of the double-ribbon flares is primarily based on the morphology of flare ribbons observed in the AIA UV passbands (1600~{\AA} and, when necessary, 1700~{\AA}). This is further substantiated by the corresponding post-flare loop structures seen in AIA EUV images and the magnetic configuration derived from HMI magnetograms.
Figure~\ref{fig:example} presents a representative event, including its HXR light curves (Figure~\ref{fig:example}a), a reconstructed HXR image in the 25–50 keV energy band (Figure~\ref{fig:example}b), and the double-ribbon structure observed in AIA 1600~{\AA} during the impulsive phase (Figure~\ref{fig:example}c). Whenever the flaring region is saturated in 1600~{\AA}, we turned to 1700~{\AA} images. Comparing UV flare ribbons with photospheric magnetograms, we excluded events with complex magnetic configurations, i.e., those exhibiting multi-ribbons or remote ribbons. Figure~\ref{fig:example}(d \& e) show the corresponding photospheric $B_z$ and $J_z$ maps for the representative event. The selection of events is further limited to those located within $45^\circ$ (Figure~\ref{fig:flare_loc}) from the central meridian to mitigate the projection effects in footpoint locations as well as in magnetic field measurements. 

\begin{figure}[ht]
  \centering
  \includegraphics[scale=0.6]{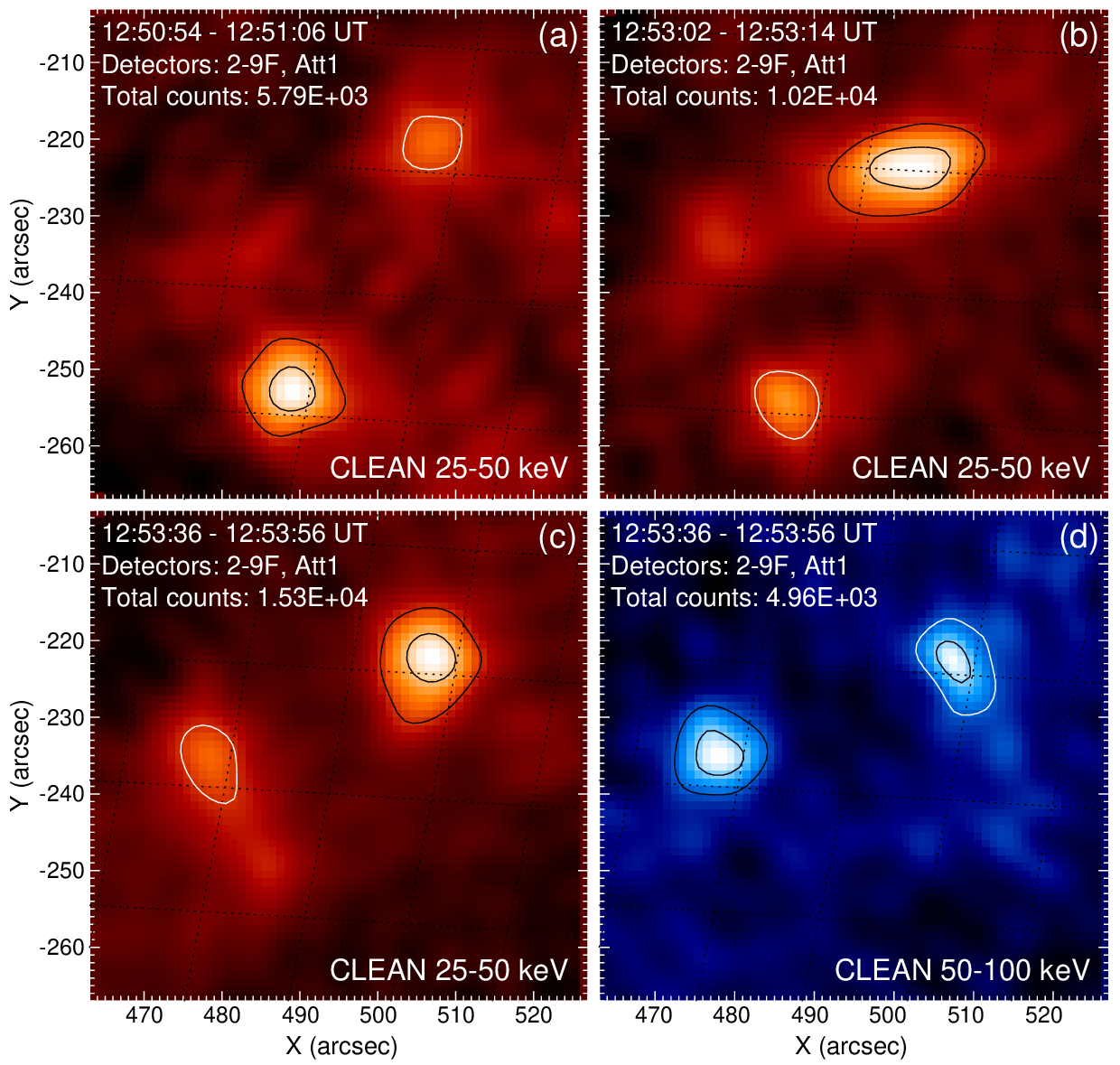}
  \caption{HXR footpoint asymmetry with time and energy dependence. The M7.3 class double-ribbon flare was observed on 2014 April 18 in NOAA AR 12036. The contours at the 50\% and 85\% of the image maximum show HXR footpoints at three time intervals and two energy bands. Between the conjugate HXR footpoints, the stronger (weaker) one is marked by black (white) contours. 
  (a--b) In the same energy band, HXR asymmetry may exhibit a reversed nature for different intervals.
  (b--c) HXR asymmetry varies when the footpoint locations change with time.
  (c--d) In the same time interval HXR asymmetry varies with different energy bands.}
  \label{fig:time-energy-depend} 
\end{figure}

\begin{figure}[ht]
  \centering
  \includegraphics[scale=0.6]{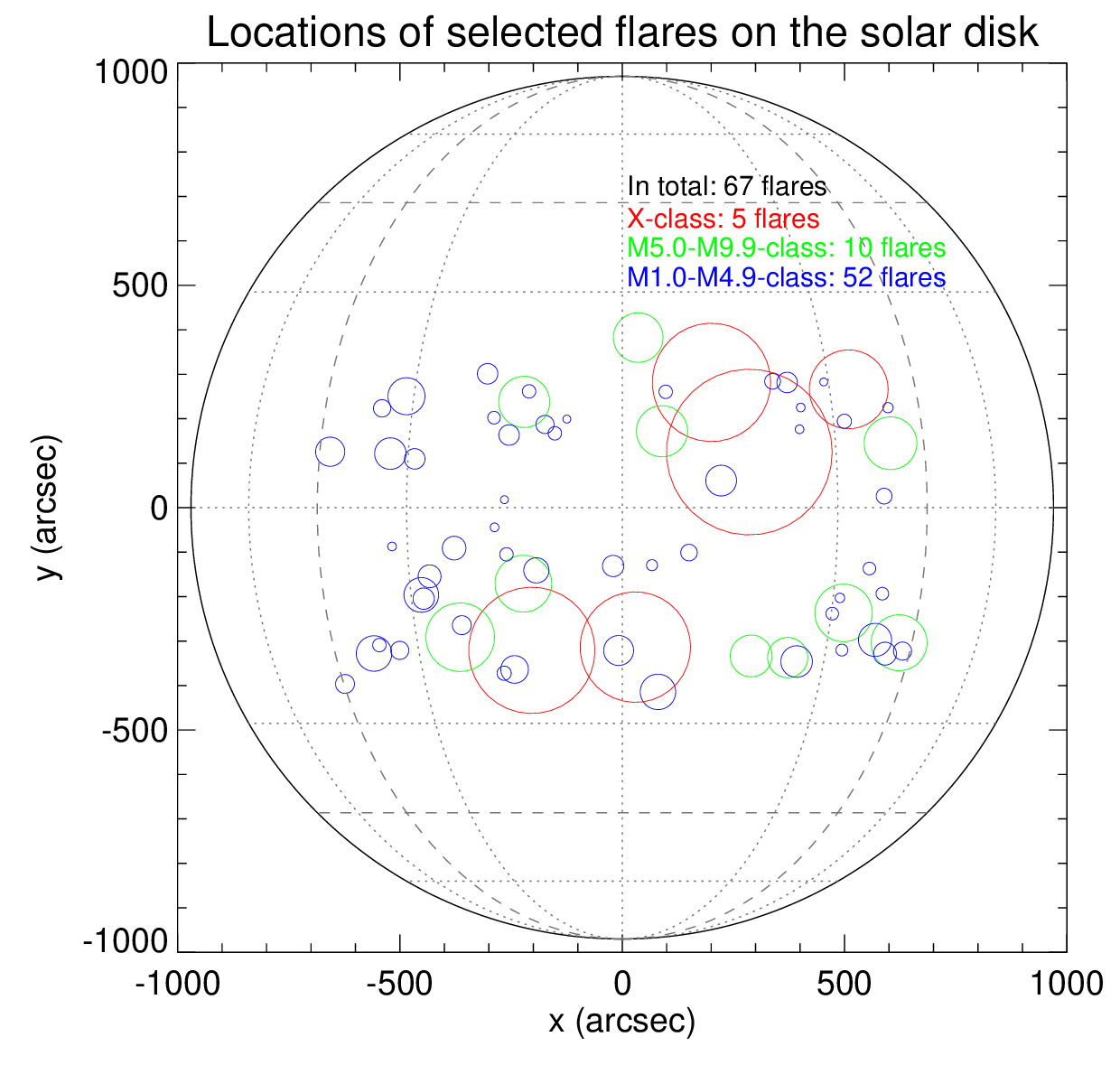}
  \caption{Heliocentric positions for the 67 flares selected in this study. X-class flares are marked by red circles, M5.0--M9.9 subgroup by light green, and M1.0--M4.9 subgroup by blue colours. The size of circles represents relative flare magnitude. Longitudes and latitudes at $\pm45^\circ$ are indicated by dashed lines.
}
  \label{fig:flare_loc} 
\end{figure}

Whenever RHESSI data are available for the selected double-ribbon flares, we reconstructed HXR images at energy ranges above 25 keV, where the nonthermal emission from the flare footpoints generally dominates over the thermal emission in the corona. A flare may have several HXR peaks, and for each peak, we have reconstructed an HXR image. However, we have selected only those time intervals that exhibit a pair of conjugate footpoint sources corresponding to the two UV flare ribbons for further analysis. Following this criterion, we have identified a total of 71 GOES M1.0-class-and-above flares with a clear double-ribbon structure. There can be multiple HXR bursts within the same flare. For example, we selected six time intervals in the 25-50 keV energy band from the M7.3 class flare observed on 2014 April 18 in NOAA AR 12036, (No. 66--71 in Table~\ref{tab:events}). The HXR images in Figure \ref{fig:time-energy-depend} show three of the six time intervals of the event, illustrating the time- and energy-dependent pattern of HXR asymmetry. 

We reconstructed 25--50 keV HXR images for 103 time intervals (Table \ref{tab:events}; more details in \S\ref{subsec:FP_indentify}) from the 67 of 71 flares, whose locations on the solar disk are shown in Figure~\ref{fig:flare_loc}. For the remaining four events, no clear double HXR sources can be reconstructed in this energy band.
For higher energy bands, we were able to obtain 50--100 keV images for 36 time intervals and 100--300 keV images for 9 time intervals from 22 flares. Due to the small sample size for higher energy bands, we have made an exhaustive investigation primarily for the 25--50 keV energy band in this study.

\subsection{HXR Footpoint Reconstruction} \label{subsec:FP_indentify}

From various image reconstruction algorithms \citep{Hurford2002, 2004SoPh..219..149A}, we have employed the CLEAN method to reconstruct HXR footpoint sources at 25--50 keV energy band with a pixel size of $1''$. CLEAN is a robust, photometrically accurate, and computationally efficient algorithm.  We have also made attempts to reconstruct 50--100 keV and 100--300 keV images. Due to lower photon counts, the quality of the images from these higher energy bands is quite limited. We selected the front segment of a detector whenever a clear source was present in its back-projection map (see Table~\ref{tab:events} for details). Usually, detectors 3--8 follow this criterion. 

Time intervals for HXR image reconstruction are selected around the peak times of individual HXR bursts to achieve better photon statistics; care has been taken to avoid attenuator state changes, satellite nighttime, South Atlantic Anomaly passage, and other complexities such as photon pileup and data decimation due to instrumental over-flooding. An important parameter to gauge the quality of reconstruction is the total number of photons in each HXR image. We adopted the rule of thumb of having a minimum of 2000 photon counts for a reliable single-image reconstruction with two sources \cite[]{Saint-Hilaire2008}. The duration of the selected time intervals is strictly a multiple of RHESSI's spin period of $\approx$ 4~s. A trial-and-error approach is also employed until sufficient photon counts are accumulated. The only exception to the aforesaid photon accumulation criterion is the 2011-August-3 M1.1-class flare (time interval \#4 in Table~\ref{tab:events}); this small burst in 25--50 keV is chosen because of the clear presence of double footpoint sources, despite the accumulation of 1390 counts only.   

\begin{ThreePartTable}
\footnotesize\setlength{\tabcolsep}{2.6pt}
    \sisetup{table-format=-1.4}
\begin{TableNotes}
  \item[\bfseries Note: ] 
  \item 5th--7th column -- GOES flare start, peak, and end times have been acquired from the Heliophysics Event Catalog (HEC) maintained by the INAF-Trieste Astronomical Observatory \citep{2017ApJ...845...49K};
  \item `Collimators' column -- RHESSI collimator configuration and attenuator state for HXR source reconstruction; ``ex'' means "excluded";
\end{TableNotes}

\begin{longtable}{|c|c|c|c|c|c|c|c|c|c|}
\caption{List of selected flare intervals for the 25-50 keV energy band.}\label{tab:events} \\ 
\toprule
\textbf{No.}          & \textbf{Date} & \textbf{NOAA AR} & \textbf{Class} & \textbf{Start} & \textbf{Peak} & \textbf{End} & \textbf{Interval (Duration [s])} & \textbf{Collimators} & \textbf{\# of Photons} \\ 
\textbf          & &  &  & \textbf{(UT)} & \textbf{(UT)} & \textbf{(UT)} &  & &  \\ \hline
\endfirsthead
\endhead
\bottomrule \addlinespace
\insertTableNotes
\endlastfoot
\toprule

\textbf{1}   & 20110216      & 11158            & M1.6           & 14:19             & 14:25            & 14:29           & 14:23:36 - 14:23:48 (12)                 & 1-9F, Att1           & 9320           \\
\textbf{2}   & 20110309      & 11166            & M1.7           & 13:17             & 14:02            & 14:13           & 13:57:52 - 13:58:16 (24)                 & 2-9F, Att1           & 2240           \\
\textbf{3}   & 20110309      & 11166            & X1.5           & 23:13             & 23:23            & 23:33           & 23:20:28 - 23:20:44 (16)                 & 1-9F, Att3           & 33600          \\
\textbf{4}   & 20110803      & 11261            & M1.1           & 03:08             & 03:37            & 03:51           & 03:34:48 - 03:34:56 (8)                 & 2-9F, Att1           & 1390           \\
\textbf{5}   & 20110803      & 11263            & M1.7           & 04:29             & 04:32            & 04:35           & 04:31:16 - 04:31:28 (12)                 & 1-9F, Att1           & 30300          \\
\textbf{6}   & 20110906      & 11283            & X2.1           & 22:12             & 22:20            & 22:40           & 22:18:20 - 22:18:28 (8)                  & 1-9F, Att3           & 96300          \\
\textbf{7}   & 20110908      & 11283            & M6.7           & 15:32             & 15:46            & 15:52           & 15:43:02 - 15:43:18 (16)                 & 1-9F, Att3           & 12000          \\
\textbf{8}   & 20110925      & 11302            & M3.7           & 15:26             & 15:33            & 15:38           & 15:30:36 - 15:30:48 (12)                 & 2-9Fex3, Att1        & 4920           \\
\textbf{9}   &               &                  &                &                   &                  &                 & 15:30:58 - 15:31:10 (12)                 & 2-9Fex3, Att1        & 6070           \\
\textbf{10}  & 20110926      & 11302            & M4.0           & 05:06             & 05:08            & 05:30           & 05:06:26 - 05:06:42 (16)                 & 1-9Fex478, Att1      & 42400          \\
\textbf{11}  & 20110926      & 11302            & M2.6           & 14:37             & 14:46            & 15:02           & 14:40:40 - 14:40:52 (12)                 & 1-9Fex5-6, Att1      & 4400           \\
\textbf{12}  & 20111002      & 11305            & M3.9           & 00:37             & 00:50            & 00:59           & 00:43:28 - 00:43:40 (12)                 & 2-9Fex3, Att1        & 6350           \\
\textbf{13}  & 20111225      & 11387            & M4.0           & 18:11             & 18:16            & 18:40           & 18:18:54 - 18:19:18 (24)                 & 1-9F, Att1           & 3350           \\
\textbf{14}  & 20111226      & 11387            & M1.5           & 02:13             & 02:27            & 02:36           & 02:21:00 - 02:21:32  (32)                & 1-9Fex2, Att1        & 4270           \\
\textbf{15}  & 20111226      & 11387            & M2.3           & 20:12             & 20:30            & 20:36           & 20:16:36 - 20:16:48 (12)                 & 1-9Fex8, Att1        & 10200          \\
\textbf{16}  & 20111231      & 11389            & M2.4           & 13:09             & 13:15            & 13:19           & 13:13:02 - 13:13:14 (12)                 & 2-9Fex8, Att1        & 7320           \\
\textbf{17}  & 20120309      & 11429            & M6.3           & 03:22             & 03:53            & 04:18           & 03:39:38 - 03:39:58 (20)                 & 3-9Fex6, Att1        & 47700          \\
\textbf{18}  &               &                  &                &                   &                  &                 & 03:40:08 - 03:40:24 (16)                 & 3-9Fex6, Att1        & 36500          \\
\textbf{19}  &               &                  &                &                   &                  &                 & 03:43:48 - 03:44:08 (20)                 & 3-9Fex6, Att3        & 12900          \\
\textbf{20}  &               &                  &                &                   &                  &                 & 03:46:12 - 03:46:24 (12)                 & 3-9F, Att3           & 5350           \\
\textbf{21}  &               &                  &                &                   &                  &                 & 04:00:00 - 04:00:20 (20)                 & 2-9F, Att3           & 14500          \\
\textbf{22}  & 20120427      & 11466            & M1.0           & 08:15             & 08:24            & 08:29           & 08:20:00 - 08:20:20 (20)                 & 1-9F, Att1           & 5650           \\
\textbf{23}  & 20120509      & 11476            & M4.7           & 12:21             & 12:32            & 12:36           & 12:31:28 - 12:31:40 (12)                 & 1-9F, Att3           & 7650           \\
\textbf{24}  & 20120510      & 11476            & M1.7           & 20:20             & 20:26            & 20:30           & 20:23:20 - 20:23:32 (12)                 & 1-9F, Att1           & 25700          \\
\textbf{25}  &               &                  &                &                   &                  &                 & 20:23:44 - 20:23:56 (12)                 & 2-9F, Att1           & 10800          \\
\textbf{26}  &               &                  &                &                   &                  &                 & 20:24:28 - 20:24:40 (12)                 & 2-9F, Att1           & 10500          \\
\textbf{27}  &               &                  &                &                   &                  &                 & 20:26:00 - 20:26:12 (12)                 & 1-9Fex2, Att1        & 25000          \\
\textbf{28}  & 20120629      & 11513            & M2.2           & 09:13             & 09:20            & 09:32           & 09:18:10 - 09:18:22 (12)                 & 1-9Fex2, Att1        & 5410           \\
\textbf{29}  & 20120630      & 11513            & M1.6           & 18:26             & 18:32            & 18:55           & 18:30:56 - 18:31:08 (12)                 & 1-9F, Att1           & 3270           \\
\textbf{30}  & 20120702      & 11515            & M3.8           & 19:59             & 20:07            & 20:13           & 20:01:48 - 20:02:08 (20)                 & 3-9F, Att1           & 27300          \\
\textbf{31}  & 20120704      & 11515            & M5.3           & 09:47             & 09:55            & 10:16           & 09:54:48 - 09:54:56 (8)                  & 1-9F, Att3           & 14400          \\
\textbf{32}  &               &                  &                &                   &                  &                 & 09:55:00 - 09:55:12 (12)                 & 1-9F, Att3           & 12300          \\
\textbf{33}  &               &                  &                &                   &                  &                 & 09:55:24 - 09:55:36 (12)                 & 1-9F, Att3           & 6310           \\
\textbf{34}  & 20120704      & 11513            & M1.8           & 16:33             & 16:39            & 16:48           & 16:35:14 - 16:35:38 (24)                 & 2-9F, Att1           & 9560           \\
\textbf{35}  & 20120706      & 11515            & M2.9           & 01:37             & 01:40            & 01:42           & 01:38:46 - 01:38:58 (12)                 & 1-9Fex3-5, Att1      & 45700          \\
\textbf{36}  & 20120712      & 11520            & X1.4           & 15:37             & 16:49            & 17:30           & 16:49:42 - 16:50:02 (20)                 & 3-9Fex4, Att3        & 7640           \\
\textbf{37}  & 20130411      & 11719            & M6.5           & 06:55             & 07:16            & 07:29           & 06:59:56 - 07:00:08 (12)                 & 2-9F, Att1           & 9550           \\
\textbf{38}  &               &                  &                &                   &                  &                 & 07:00:24 - 07:00:36 (12)                 & 2-9F, Att1           & 7070           \\
\textbf{39}  &               &                  &                &                   &                  &                 & 07:02:24 - 07:02:36 (12)                 & 2-9F, Att1           & 6620           \\
\textbf{40}  &               &                  &                &                   &                  &                 & 07:02:36 - 07:02:48 (12)                 & 2-9F, Att1           & 8420           \\
\textbf{41}  &               &                  &                &                   &                  &                 & 07:02:48 - 07:03:00 (12)                 & 2-9F, Att1           & 8070           \\
\textbf{42}  &               &                  &                &                   &                  &                 & 07:03:14 - 07:03:26 (12)                 & 2-9F, Att1           & 6920           \\
\textbf{43}  & 20130502      & 11731            & M1.1           & 04:58             & 05:10            & 05:19           & 05:04:40 - 05:05:00 (20)                 & 2-9F, Att1           & 25100          \\
\textbf{44}  & 20130503      & 11731            & M1.3           & 16:39             & 16:55            & 17:22           & 16:45:48 - 16:46:04 (16)                 & 2-9F, Att1           & 8490           \\
\textbf{45}  & 20131022      & 11875            & M1.0           & 00:14             & 00:22            & 00:28           & 00:18:00 - 00:18:20 (20)                 & 1-9F, Att1           & 9070           \\
\textbf{46}  & 20131028      & 11882            & M2.7           & 14:46             & 15:01            & 15:04           & 15:00:00 - 15:00:24 (24)                 & 1-9Fex4, Att1        & 9450           \\
\textbf{47}  & 20131028      & 11882            & M4.4           & 15:07             & 15:15            & 15:21           & 15:10:28 - 15:10:56 (28)                 & 2-9F, Att1           & 38300          \\
\textbf{48}  & 20131107      & 11890            & M2.4           & 14:15             & 14:25            & 14:31           & 14:29:00 - 14:29:20 (20)                 & 1-6F, Att1           & 15900          \\
\textbf{49}  & 20140107      & 11946            & M1.0           & 03:49             & 03:53            & 03:56           & 03:51:08 - 03:51:20 (12)                 & 1-9F, Att1           & 6120           \\
\textbf{50}  & 20140107      & 11944            & M7.2           & 10:07             & 10:13            & 10:37           & 10:10:35 - 10:10:43 (8)                 & 1-9Fex3, Att0        & 4970           \\
\textbf{51}  &               &                  &                &                   &                  &                 & 10:10:58 - 10:11:18 (20)                 & 2-9F, Att1           & 141000         \\
\textbf{52}  &               &                  &                &                   &                  &                 & 10:11:28 - 10:11:48 (20)                 & 2-9F, Att3           & 181000         \\
\textbf{53}  &               &                  &                &                   &                  &                 & 10:11:48 - 10:12:08 (20)                 & 2-9F, Att3           & 127000         \\
\textbf{54}  &               &                  &                &                   &                  &                 & 10:12:08 - 10:12:28 (20)                 & 2-9F, Att3           & 69000          \\
\textbf{55}  & 20140201      & 11967            & M3.0           & 07:14             & 07:23            & 07:36           & 07:18:40 - 07:19:00 (20)                 & 1-9Fex3, Att1        & 6170           \\
\textbf{56}  & 20140202      & 11968            & M2.6           & 06:24             & 06:34            & 06:53           & 06:33:52 - 06:34:08 (16)                 & 1-9F, Att1           & 12900          \\
\textbf{57}  & 20140211      & 11974            & M1.7           & 03:22             & 03:31            & 03:39           & 03:27:20 - 03:27:44 (24)                 & 1-9F, Att1           & 13900          \\
\textbf{58}  & 20140214      & 11974            & M1.6           & 12:29             & 12:40            & 12:45           & 12:34:48 - 12:35:28 (40)                 & 3-9Fex4, Att1        & 5680           \\
\textbf{59}  & 20140328      & 12017            & M2.0           & 19:04             & 19:18            & 19:27           & 19:18:08 - 19:18:24 (16)                 & 2-9F, Att1           & 2770           \\
\textbf{60}  & 20140328      & 12017            & M2.6           & 23:44             & 23:51            & 23:58           & 23:49:02 - 23:49:18 (16)                 & 2-9F, Att1           & 8720           \\
\textbf{61}  & 20140329      & 12017            & X1.0           & 17:35             & 17:48            & 18:04           & 17:45:32 - 17:45:48 (16)                 & 1-9Fex8, Att3        & 53100          \\
\textbf{62}  &               &                  &                &                   &                  &                 & 17:46:08 - 17:46:24 (16)                 & 1-9Fex8, Att3        & 35600          \\
\textbf{63}  &               &                  &                &                   &                  &                 & 17:46:25 - 17:46:41 (16)                 & 1-9Fex8, Att3        & 42200          \\
\textbf{64}  &               &                  &                &                   &                  &                 & 17:46:42 - 17:46:58 (16)                 & 1-9Fex8, Att3        & 37700          \\
\textbf{65}  &               &                  &                &                   &                  &                 & 17:46:58 - 17:47:14 (16)                 & 1-9Fex8, Att3        & 35900          \\
\textbf{66}  & 20140418      & 12036            & M7.3           & 12:31             & 13:03            & 13:20           & 12:50:42 - 12:50:54 (12)                 & 2-9F, Att1           & 5100           \\
\textbf{67}  &               &                  &                &                   &                  &                 & 12:50:54 - 12:51:06 (12)                 & 2-9F, Att1           & 5790           \\
\textbf{68}  &               &                  &                &                   &                  &                 & 12:53:02 - 12:53:14 (12)                 & 2-9F, Att1           & 10200          \\
\textbf{69}  &               &                  &                &                   &                  &                 & 12:53:36 - 12:53:56 (20)                 & 2-9F, Att1           & 15300           \\
\textbf{70}  &               &                  &                &                   &                  &                 & 12:54:02 - 12:54:22 (20)                 & 2-9F, Att1           & 19100          \\
\textbf{71}  &               &                  &                &                   &                  &                 & 12:58:20 - 12:58:32 (12)                 & 2-9F, Att1           & 5130           \\
\textbf{72}  & 20140825      & 12146            & M2.0           & 14:46             & 15:11            & 15:31           & 15:00:42 - 15:00:58 (16)                 & 2-9F, Att1           & 6660           \\
\textbf{73}  & 20140923      & 12172            & M2.3           & 23:03             & 23:16            & 23:28           & 23:09:12 - 23:09:36 (24)                 & 3-9F, Att1           & 29600          \\
\textbf{74}  & 20140928      & 12173            & M5.1           & 02:39             & 02:58            & 03:19           & 02:42:43 - 02:43:07 (24)                 & 2-9Fex3, Att1        & 10700          \\
\textbf{75}  &               &                  &                &                   &                  &                 & 02:45:13 - 02:45:29 (16)                 & 2-9Fex3, Att1        & 8000           \\
\textbf{76}  &               &                  &                &                   &                  &                 & 02:46:36 - 02:46:48 (12)                 & 2-9Fex3, Att1        & 13000          \\
\textbf{77}  & 20141020      & 12192            & M4.5           & 16:00             & 16:37            & 16:55           & 16:08:32 - 16:08:52 (20)                 & 2-9F, Att1           & 26600          \\
\textbf{78}  & 20141020      & 12192            & M1.7           & 19:53             & 20:04            & 20:13           & 19:57:16 - 19:57:28 (12)                 & 3-9F, Att1           & 26300          \\
\textbf{79}  & 20141022      & 12192            & M8.7           & 01:16             & 01:59            & 02:28           & 01:38:26 - 01:38:34 (8)                  & 2-9F, Att1           & 62100          \\
\textbf{80}  &               &                  &                &                   &                  &                 & 01:39:02 - 01:39:10 (8)                  & 2-9F, Att1           & 59500          \\
\textbf{81}  &               &                  &                &                   &                  &                 & 01:39:20 - 01:39:28 (8)                  & 2-9F, Att3           & 67400          \\
\textbf{82}  & 20141022      & 12192            & X1.6           & 14:02             & 14:28            & 14:50           & 14:05:14 - 14:05:26 (12)                 & 1-9F, Att3           & 97200          \\
\textbf{83}  &               &                  &                &                   &                  &                 & 14:05:35 - 14:05:47 (12)                 & 1-9F, Att3           & 183000         \\
\textbf{84}  &               &                  &                &                   &                  &                 & 14:06:17 - 14:06:29 (12)                 & 1-9F, Att5           & 324000         \\
\textbf{85}  & 20141024      & 12192            & M4.0           & 07:37             & 07:48            & 08:06           & 07:40:24 - 07:40:36 (12)                 & 1-9F, Att1           & 45100          \\
\textbf{86}  & 20141026      & 12192            & M4.2           & 18:07             & 18:15            & 18:20           & 18:08:28 - 18:08:40 (12)                 & 3-9F, Att1           & 41200          \\
\textbf{87}  & 20141027      & 12192            & M7.1           & 00:06             & 00:34            & 00:44           & 00:19:28 - 00:19:40 (12)                 & 1-9Fex3-4, Att3      & 7620           \\
\textbf{88}  &               &                  &                &                   &                  &                 & 00:21:52 - 00:22:04 (12)                 & 1-9Fex3, Att3        & 4100           \\
\textbf{89}  & 20141109      & 12205            & M2.3           & 15:24             & 15:32            & 15:38           & 15:30:08 - 15:30:20 (12)                 & 1-9F, Att1           & 17600          \\
\textbf{90}  & 20141201      & 12222            & M1.8           & 06:26             & 06:41            & 06:59           & 06:32:44 - 06:33:16 (32)                 & 3-9Fex5, Att1        & 6700           \\
\textbf{91}  & 20150103      & 12253            & M1.1           & 09:40             & 09:47            & 09:50           & 09:45:20 - 09:45:36 (16)                 & 1-9F, Att1           & 16400          \\
\textbf{92}  & 20150126      & 12268            & M1.1           & 16:46             & 16:53            & 16:58           & 16:50:00 - 16:50:20 (20)                 & 1-9Fex245, Att1      & 3410           \\
\textbf{93}  & 20150129      & 12268            & M2.1           & 11:32             & 11:42            & 11:52           & 11:35:40 - 11:35:52 (12)                 & 1-9F, Att1           & 11700          \\
\textbf{94}  & 20150310      & 12297            & M2.9           & 23:46             & 00:02            & 00:06           & 00:00:00 - 00:00:12 (12)                 & 2-9F, Att1           & 105000         \\
\textbf{95}  & 20150312      & 12297            & M3.2           & 04:41             & 04:46            & 04:50           & 04:43:20 - 04:43:32 (12)                 & 1-9F, Att1           & 106000         \\
\textbf{96}  & 20150312      & 12297            & M2.7           & 21:44             & 21:51            & 21:56           & 21:47:48 - 21:48:00 (12)                 & 2-9F, Att1           & 32700          \\
\textbf{97}  & 20150315      & 12297            & M1.2           & 22:42             & 23:22            & 23:38           & 22:45:44 - 22:45:56 (12)                 & 1-9F, Att1           & 20900          \\
\textbf{98}  & 20150316      & 12297            & M1.6           & 10:39             & 10:58            & 11:17           & 10:51:08 - 10:51:32 (24)                 & 1-9F, Att1           & 2620           \\
\textbf{99}  & 20150408      & 12320            & M1.4           & 14:37             & 14:43            & 14:47           & 14:40:24 - 14:40:44 (20)                 & 1-9F, Att0           & 11800          \\
\textbf{100} &               &                  &                &                   &                  &                 & 14:41:48 - 14:42:08 (20)                 & 1-9F, Att1           & 15000          \\
\textbf{101} & 20150621      & 12371            & M2.6           & 02:06             & 02:36            & 03:02           & 02:10:20 - 02:10:40 (20)                 & 2-9F, Att1           & 9370           \\
\textbf{102} & 20150622      & 12371            & M6.5           & 17:39             & 18:23            & 18:51           & 18:23:30 - 18:23:46 (16)                 & 2-9F, Att3           & 16200          \\
\textbf{103} & 20150822      & 12403            & M3.5           & 21:19             & 21:24            & 21:28           & 21:21:48 - 21:22:08 (20)                 & 1-8Fex2, Att3        & 56400         
\end{longtable}
\end{ThreePartTable}

\subsection{Mapping of Photospheric Parameters} \label{subsec:PVEC}
To obtain the magnetic fields as well as electric currents at the HXR footpoints, the reconstructed RHESSI images (e.g., Figure \ref{fig:example}b) are remapped to a Cylindrical Equal Area (CEA) Cartesian coordinate system centered on the tracked active region, same as those provided by the hmi.sharp\_cea\_720s data series of the Spaceweather HMI Active Region Patch (SHARP), which have a pixel size of 0.36 Mm and a cadence of 12 min \citep{2014SoPh..289.3483H}. The HXR sources are then superimposed onto the vertical component $B_{z}$ maps of vector magnetograms (e.g., Figure \ref{fig:example}d) and the corresponding PVEC $J_{z}$ maps (e.g., Figure \ref{fig:example}e).

For each flare, we worked with HMI vector magnetograms taken immediately before the flare onset (e.g., Figure~\ref{fig:example}d), as polarimetric measurements can be affected by electron beams during the flare impulsive phase. The vector magnetograms allow the calculation of PVECs through Amp\`{e}re's law,
\begin{equation} 
    \mu_0 J_z = \frac{\partial B_y}{\partial x} - \frac{\partial B_x}{\partial y}.
    \label{eq:jz} 
\end{equation}
The computation of $J_z$ is subject to the 180{\degree} ambiguity and the large intrinsic uncertainties involved in measuring the transverse components of photospheric magnetic fields \citep[$\sim\,$100 G;][]{2014SoPh..289.3483H}. These uncertainties can be further amplified by numerical differences to implement the partial derivatives in Eq.~\ref{eq:jz}. \citealt{He2020} estimated that the uncertainty of $J_z$ is about 22 mA~m$^{-2}$ for AR magnetic fields; as a comparison, the average $|J_z|$ in quiet-Sun regions is normally below 10 mA~m$^{-2}$. To reduce the random noise resulting from these uncertainties, we obtained a time-averaged $J_z$ map from five $J_z$ maps derived from five pre-flare vector magnetograms acquired during an hour before the flare onset. In our statistical analysis, the standard deviation derived from these five $J_z$ maps is taken as the uncertainty. Figure \ref{fig:example}e shows the $J_z$ map of an exemplary event, with a pixel size of $\approx370 $ km. One may note that each of the two HXR footpoints is associated with a strong $J_z$ ribbon at the opposite sides of the PIL.

\section{Statistical results} \label{sec:statistics}

\subsection{Quantification of HXR footpoint asymmetry} \label{subsec:HXR_FP_asymmrtry}

\begin{figure}
    \centering 
    \includegraphics[width=0.9\linewidth]{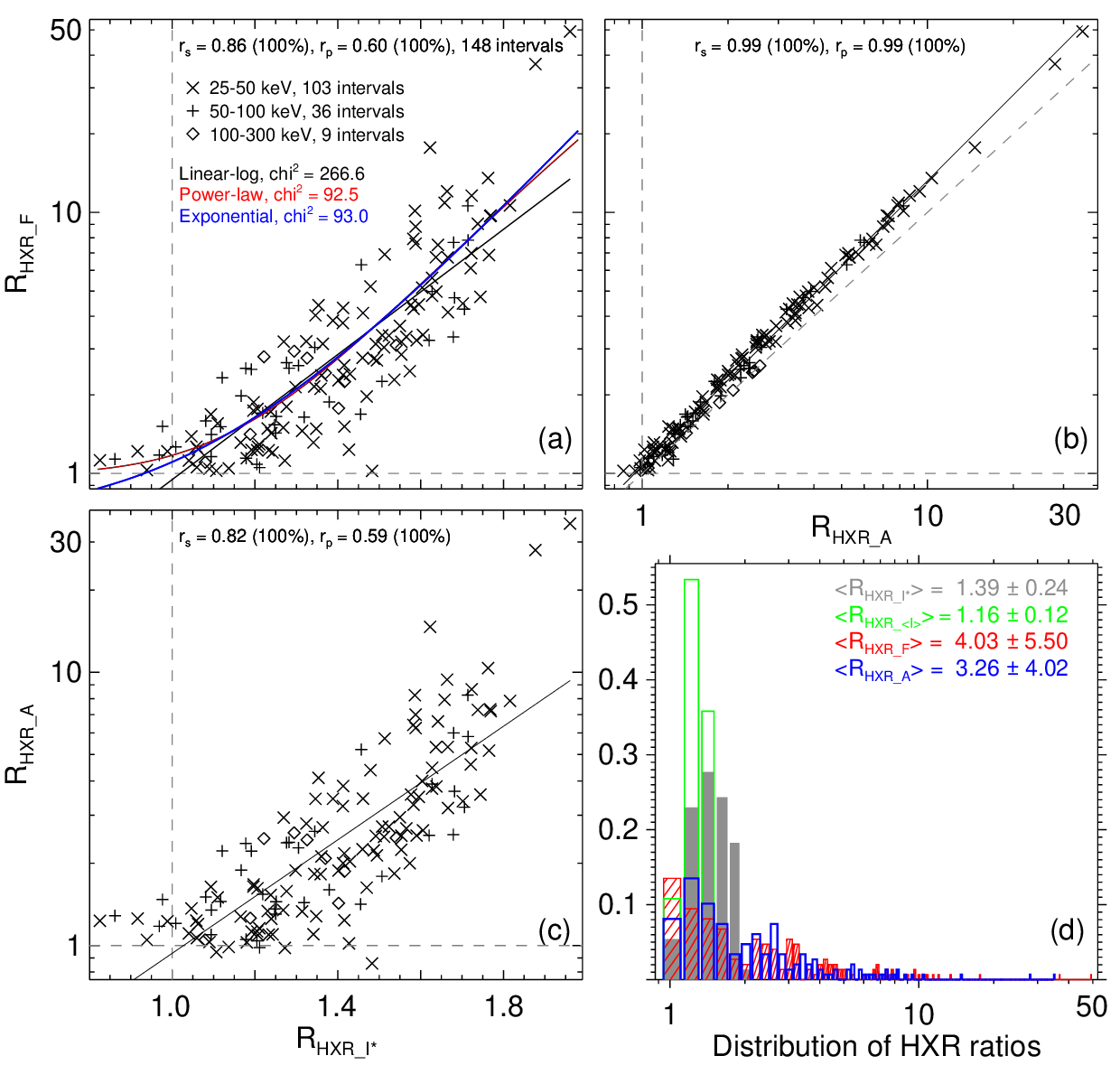}
    \caption{Scatter plots of $R_\text{HXR\_I*}$, $R_\mathrm{HXR\_F}$ and $R_\mathrm{HXR\_A}$ for three energy bands: 25-50 keV (`$\times$' symbols), 50-100 keV (`+' symbols) and 100-300 keV (`$\diamond$' symbols). The horizontal dashed lines in (a) and (b) demonstrates that no data below $R_\mathrm{HXR\_F}=1$. The vertical dashed lines in (a) and (c) mark the unity of $R_\text{HXR\_I*}$, and that in (b) marks the unity of $R_\mathrm{HXR\_A}$. The diagonal dashed line in (b) marks $R_\mathrm{HXR\_F}$ = $R_\mathrm{HXR\_A}$.} The black solid lines represent a linear fit in linear-logarithm coordinates in (a) and (c) and in linear coordinates in (b), respectively. The power-law (red), and exponential (blue) fits of the scatter plot in (a) is obtained after excluding the two outlier data points in the upper-right corner. The distributions of the HXR asymmetry ratios, including the mean intensity ratio $R_{\mathrm{HXR\_\langle I\rangle}}$, along with their mean values, are shown in (d).
    \label{fig:HXRF-HXRI}
        
\end{figure}

To quantify the asymmetry between the two conjugate footpoints in a flare, we first determine the HXR source size by the contour at the level of 50\% maximum intensity in the reconstructed CLEAN images. We define the asymmetry ratio in terms of the HXR flux as follows,
\begin{equation}
    R_{\text{HXR\_F}} = \frac{F_{\text{HXR,1}}}{F_{\text{HXR,2}}} = \frac{\int_{A_1} I_{\text{HXR,1}} \,dA}{\int_{A_2} I_{\text{HXR,2}}\,dA}  \label{eq:ratio_hxr_f}.
\end{equation}
Hereafter, the subscripts 1 and 2 refer to the stronger and weaker HXR footpoint, respectively, as determined by the HXR flux of FP1 ($F_\mathrm{HXR,1}$) and FP2 ($F_\mathrm{HXR,2}$), which is obtained by integrating the HXR intensity $I_{\text{HXR,i}}$ in units of photons~$\mathrm{s}^{-1}\,\mathrm{cm}^{-2}\,\mathrm{sr}^{-1}$ over each individual source area $A_i$ ($i=1,\,2$). $R_{\mathrm{HXR\_F}} = 1$ means that the two footpoints are ideally symmetric. Based on this definition, we always have $R_{\mathrm{HXR\_F}} \ge 1$. The HXR source area ratio is denoted as 
\begin{equation}
R_\mathrm{HXR\_A}=\frac{A_1}{A_2}.
\end{equation} 
We define the asymmetry ratio in terms of the footpoint maximum brightness: 
\begin{equation}
    R_{\text{HXR\_I*}} = \frac{\max(I_{\text{HXR,1}})}{\max(I_{\text{HXR,2}})}, \label{eq:ratio_hxr_imax}
\end{equation} where the numerator and denominator refer to the maximum HXR intensity of the stronger and weaker FP, respectively. The ratio of mean brightness is denoted as 
\begin{equation}
    R_{\mathrm{HXR\_{\langle I\rangle}}} = \frac{\langle I_{\text{HXR,1}}\rangle}{\langle I_{\text{HXR,2}}\rangle}.\label{eq:ratio_hxr_imean}
\end{equation}
Since the source strength is defined by flux rather than by maximum or mean brightness, we have a few exceptions where the brightness ratio does not exceed unity (Figure~\ref{fig:HXRF-HXRI}). Similarly, hereafter we define the ratio of a physical variable, say, $V$ at the conjugate HXR FPs as follows, 
\begin{equation}
    R_V = \frac{V_1}{V_2} = \frac{\text{parameter V within FP1}}{\text{parameter V within FP2}} \label{eq:ratio}, 
\end{equation}
where the parameter $V$ is sampled within the 50\% contour level of each individual HXR source. 

The uncertainties in HXR source quantities (photon flux, maximum intensity, and source area) arise primarily from instrumental effects and image reconstruction using the CLEAN algorithm, which are difficult to quantify. As described in \S\ref{subsec:FP_indentify}, these effects are mitigated by selecting time intervals with sufficient photon counts and by excluding periods affected by artifacts. The use of ratio-based quantities further reduces systematic biases. 

To assess the statistical relationships between different physical parameters, we employ both the Spearman ($r_{\mathrm{s}}$) and Pearson ($r_{\mathrm{p}}$) correlation coefficients. The Pearson correlation measures the strength and direction of a linear relationship between two variables, assuming normally distributed data. The Spearman correlation is a non-parametric measure that evaluates monotonic relationships based on the ranked values of the data, making it less sensitive to outliers and applicable to non-linear trends. For both coefficients, we give the confidence level as a quantitative measure of the likelihood that the detected correlation is not due to random chance. As demonstrated in Figure~\ref{fig:HXRF-HXRI}(a \& c), both $R_{\mathrm{HXR\_F}}$ and $R_{\mathrm{HXR\_A}}$ show a positive monotonic relation with  $R_{\text{HXR\_I*}}$ in the linear-logarithm plots, while $R_{\mathrm{HXR\_F}}$ is strongly linearly correlated with $R_{\mathrm{HXR\_A}}$ (Figure~\ref{fig:HXRF-HXRI}b). As a result, the ratio of mean intensity ratio $R_{\mathrm{HXR\_{\langle I\rangle}}}$ is highly concentrated around the mean value of 1.16. In contrast, $R_{\mathrm{HXR\_F}}$, $R_{\text{HXR\_I*}}$, and $R_{\mathrm{HXR\_A}}$ are highly spread and right-skewed (Figure~\ref{fig:HXRF-HXRI}d). The implications of these relations are discussed in detail in \S\ref{subsec:discuss_asymmetry}.

\subsection{Test on the magnetic mirroring effect} \label{subsec:mm}
\begin{figure}
    \centering
    \includegraphics[width=0.9\linewidth]{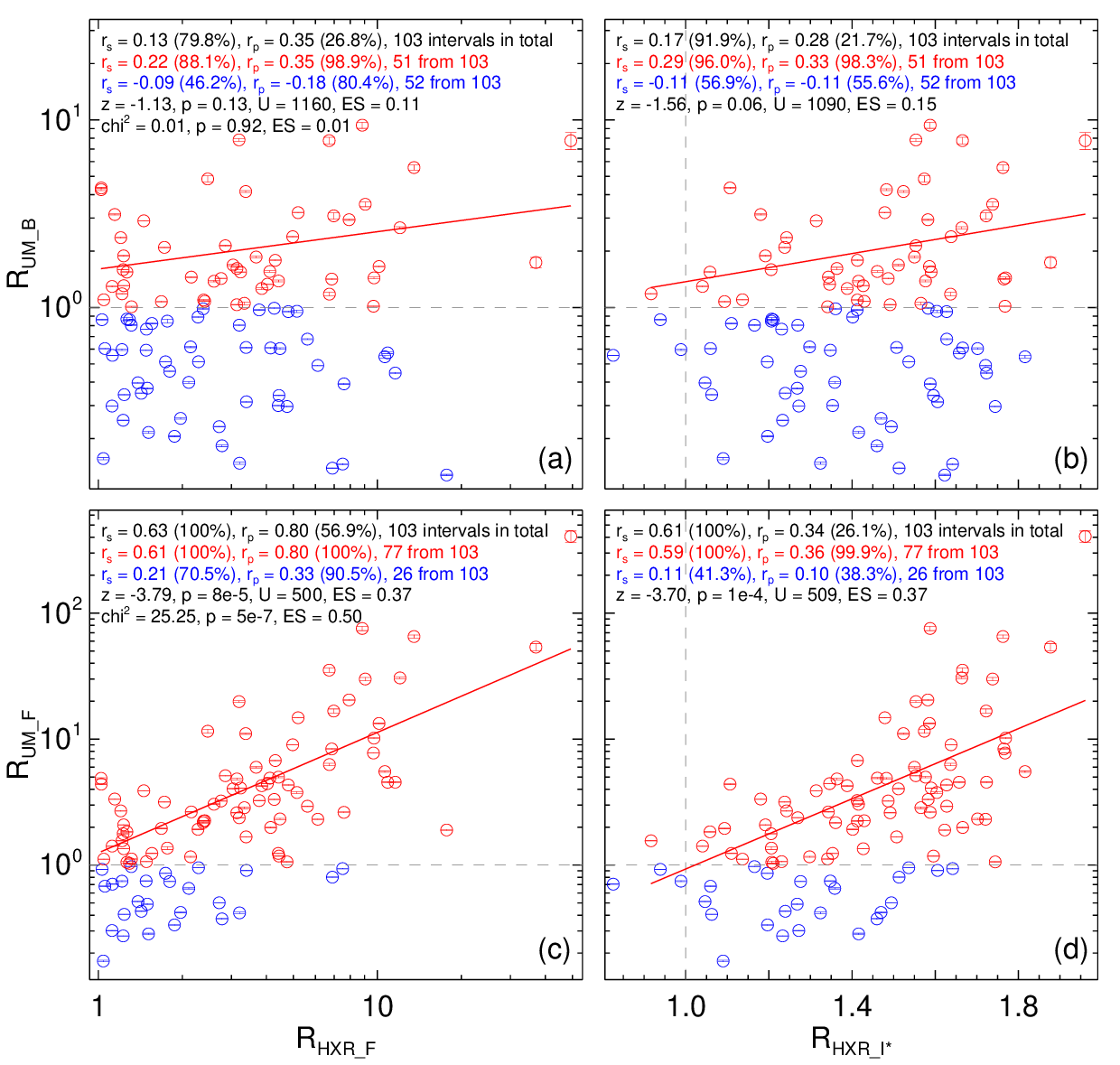}    
    \caption{Scatter plots for magnetic parameter ratios vs HXR asymmetry ratios at 25-50 keV. Anti-MM cases ($R_\mathrm{UM\_F} > 1$ or $R_\mathrm{UM\_B} > 1$) are marked in red; pro-MM cases ($R_\mathrm{UM\_F} < 1$ or $R_\mathrm{UM\_B} < 1$) are marked in blue. The horizontal dashed lines in all panels mark the unity of $R_\mathrm{UM\_F}$ or $R_\mathrm{UM\_B}$. 
    The vertical dashed lines in (b) and (d) mark the unity of $R_\text{HXR\_I*}$. The estimated errors for $R_\mathrm{UM\_B}$ and $R_\mathrm{UM\_F}$ are shown by the vertical bars. } 
 \label{fig:HXR-UM}
\end{figure}  

Previous studies on HXR footpoint asymmetry often frame their analysis in terms of magnetic field strength in the context of magnetic mirroring \cite[e.g.,][]{Sakao1996,Goff2004,Yang2012}.
Here we employ magnetic field parameters including both magnetic field strength and magnetic flux in the format of ratios to quantify the asymmetry of photospheric magnetic field at the conjugate HXR footpoints. The ratio of unsigned magnetic field strength $R_\mathrm{UM\_B}$ is defined as
\begin{equation}
    R_\mathrm{UM\_B} = \frac{B_\mathrm{UM,1}}{B_\mathrm{UM,2}}=\frac{\langle |B_{z,1}| \rangle}{\langle |B_{z,2}| \rangle}, \label{eq:ratio_umb}
\end{equation}  
where $\langle|B_z|\rangle$ denotes the mean strength of the vertical component of photospheric magnetic field over each individual HXR source. Similarly, the ratio of unsigned magnetic flux $R_\mathrm{UM\_F}$ is defined as  
\begin{equation}
    R_\mathrm{UM\_F} = \frac{F_\mathrm{UM,1}}{F_\mathrm{UM,2}}  = \frac{\int_{A_1} |B_{z,1}|\,dA}{\int_{A_2} |B_{z,2}|\,dA},  \label{eq:ratio_umf}
\end{equation}
where the integration covers the 50\% contour level of each individual HXR source. The uncertainties in $B_z$ are derived from the SHARP data products, which provide pixel-level error estimates based on the inversion of Stokes profiles (see \citealt{2014SoPh..289.3483H}).

In Figure~\ref{fig:HXR-UM} we show the scatter plots of HXR flux ratio $R_{\text{HXR\_F}}$ and maximum brightness ratio $R_{\text{HXR\_I*}}$ versus $R_{\text{UM\_F}}$ and $R_{\text{UM\_B}}$. Cases with $R_\mathrm{UM\_F} < 1$ or $R_\mathrm{UM\_B} < 1$ (traditionally S-type) may be intuitively understood to conform with the magnetic mirroring scenario (`pro-MM' hereafter) and marked in blue, while those with $R_\mathrm{UM\_F} >1$ or $R_\mathrm{UM\_B} > 1$ (N-type) are contradicting the magnetic mirroring scenario (`anti-MM' hereafter) and marked in red. Overall, we found a positive correlation between the HXR asymmetry ratio and the magnetic field ratio at the HXR footpoints, for both the anti- and pro-MM cases, and the positive correlation is particularly strong for the anti-MM cases. 

Specifically, the ratio of mean field strength $R_{\text{UM\_B}}$ is not significantly correlated with HXR flux asymmetry $R_{\text{HXR\_F}}$ nor brightness asymmetry $R_{\text{HXR\_I*}}$ (Figure~\ref{fig:HXR-UM} (a \& b)). Approximately half of the cases (50\% or 52 intervals from 103) are pro-MM ($R_{\text{UM\_B}}<1$) and exhibit a weak negative correlation; the rest of cases are anti-MM ($R_{\text{UM\_B}}>1$) and exhibit a weak positive correlation. 
The Mann-Whitney U-test (Appendix~\ref{app:U_test}) demonstrates that the two populations are not significantly different in terms of $R_{\text{HXR\_I*}}$ or $R_{\text{HXR\_F}}$. This result is similar to the study by \citet[][their Figure 3]{Goff2004} using Yohkoh data.    

In contrast, the ratio of unsigned flux $R_{\text{UM\_F}}$ (Figure~\ref{fig:HXR-UM}(c \& d)) is significantly correlated with both $R_{\text{HXR\_F}}$ and $R_{\text{HXR\_I*}}$. The majority of the cases are anti-MM (75\% or 77 intervals from 103). The Mann-Whitney U-test demonstrates that the anti- ($R_{\text{UM\_F}}>1$) and pro-MM ($R_{\text{UM\_F}}<1$) cases are significantly different in terms of both $R_{\text{HXR\_I*}}$ and $R_{\text{HXR\_F}}$. $R_{\text{UM\_F}}$ shows a strong positive correlation with $R_{\text{HXR\_F}}$ for the anti-MM cases ($R_{\text{UM\_F}}>1$) but a weak positive correlation for the pro-MM cases ($R_{\text{UM\_F}}<1$), with Spearman and Pearson correlations generally agreeing with each other. Regarding its relation to $R_{\text{HXR\_I*}}$, $R_{\text{UM\_F}}$ shows a strong Spearman but weak Pearson correlation coefficients, indicating a nonlinear positive correlation dominated by the anti-MM cases ($R_{\text{UM\_F}}>1$), whereas both correlations are weak for the pro-MM cases ($R_{\text{UM\_F}}<1$).

\subsection{Test on the electric current effect} \label{subsec:ec}
\begin{figure}
    \centering
    \includegraphics[width=0.9\linewidth]{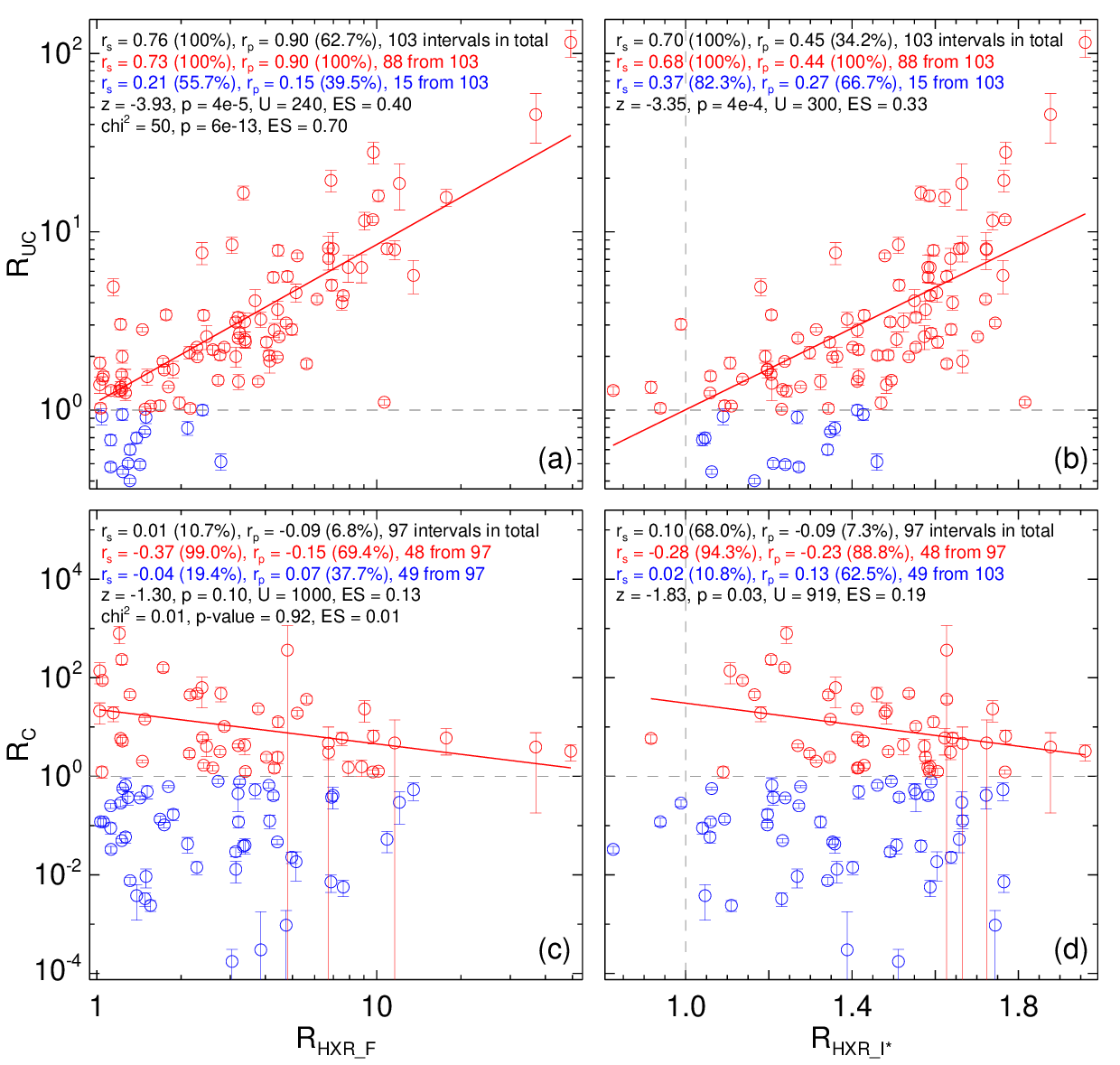}
    \caption{Scatter plots for PVEC ratios vs HXR asymmetry ratio at 25-50 keV. Intervals with $R_\mathrm{UC} > 1$ or $R_\mathrm{C} > 1$ are marked in red, while those with $R_\mathrm{UC} < 1$ or $R_\mathrm{C} < 1$ are marked in blue. The horizontal dashed lines in all panels mark the unity of $R_\mathrm{UC}$ or $R_\mathrm{C}$. The vertical dashed lines in (b) and (d) mark the unity of $R_\text{HXR\_I*}$. The estimated uncertainty of $R_\mathrm{UC}$ and $R_\mathrm{C}$ is shown in vertical error bars. $R_\mathrm{C}$ with extreme values, due to $I_z^+$ or $I_z^-$ approaching zero (6 of 103 intervals), were excluded from the statistical analysis.} 
    \label{fig:HXR-UC-EC}
\end{figure}

Electric currents may have various potential impacts on the precipitation of electrons and thereby on HXR emission \citep{Canfield1993a,2006ApJ...651..553Z,Haerendel2017}, which have not been investigated systematically. To test the effect of electric currents on the HXR asymmetry, we adopt the ratio of unsigned PVEC as defined by 
\begin{equation}
    R_{\text{UC}} = \frac{|I_\mathrm{z,1}|}{|I_\mathrm{z,2}|} = \frac{\int_{A_1} |J_{z,1}| dA}{\int_{A_2} |J_{z,2}| dA}\label{eq:ratio_uc},
\end{equation}
where the magnitude of the PVEC density, $|J_{z,i}|$ ($i=1,\,2$), is integrated over the area ($A_i$) of the HXR footpoint $i$ to obtain the unsigned PVEC $|I_{z,i}|$. Figure~\ref{fig:HXR-UC-EC} (a \& b) show the scatter plots of $R_{\text{HXR\_F}}$ and $R_{\text{HXR\_I*}}$ in relation to $R_{\text{UC}}$, respectively. The uncertainties in PVEC are detailed in \S\ref{subsec:PVEC}.

The ratio of signed PVEC is defined by 
\begin{equation}
    R_\mathrm{C} = \frac{I^+_{z,1}/I^+_{z,2}}{I^-_{z,1}/I^-_{z,2}} = 
    \frac{I^+_{z,1}}{I^+_{z,2}} \times \frac{I^-_{z,2}}{I^-_{z,1}} = 
    \frac{\int_{A_1} J^+_{z,1}\,dA}{\int_{A_2} J^+_{z,2}\,dA} \times 
    \frac{\int_{A_2} J^-_{z,2}\,dA }{\int_{A_1} J^-_{z,1}\,dA}
    \label{eq:ratio_c},
\end{equation}
where positive (negative) current density $J^+_{z,i}$ ($J^-_{z,i}$; $i=1,\,2$) is integrated over the area $A_i$ of the respective HXR footpoint to obtain the corresponding positive (negative) electric current $I^+_{z,i}$ ($I^-_{z,i}$). The PVEC density $J^+_{z,i}$ ($J^-_{z,i}$) indicates a drop (rise) in electric potential from the surface to the lower atmosphere, therefore facilitating (hindering) the electron precipitation. Thus, $R_\mathrm{C}$ gauges the effect of the electric field that is related to the PVEC density through Ohm's law $E_z=\sigma J_z$, by a ratio of $I^+_{z,1}/I^+_{z,2}$ over $I^-_{z,1}/I^-_{z,2}$. The former (latter) factor, if larger than 1, is expected to pose a positive (negative) effect on the HXR footpoint asymmetry, i.e., making $R_{\text{HXR\_F}}$ and $R_{\text{HXR\_I*}}$ larger (smaller). Essentially, a positive correlation is expected between $R_{\text{C}}$ and the HXR footpoint asymmetry. Figure~\ref{fig:HXR-UC-EC}(c \& d) shows the scatter plots of $R_{\text{HXR\_F}}$ and $R_{\text{HXR\_I*}}$ in relation to $R_{\text{C}}$, respectively. 

Both the HXR flux asymmetry ratio $R_{\text{HXR\_F}}$ and the maximum brightness asymmetry ratio $R_{\text{HXR\_I*}}$ show a strong positive correlation with the unsigned vertical current ratio $R_{\mathrm{UC}}$. The Spearman coefficients are high and statistically significant, while the Pearson coefficients are also positive though systematically smaller, indicating a predominantly monotonic but not strictly linear dependence. The Mann-Whitney U-test confirms that the $R_{\mathrm{UC}} > 1$ and $R_{\mathrm{UC}} < 1$ populations differ significantly in terms of HXR asymmetry, with medium-to-large effect sizes. 

In contrast, the signed PVEC ratio $R_{C}$ exhibits weaker and less consistent correlations with $R_{\text{HXR\_F}}$ and $R_{\text{HXR\_I*}}$. The $R_{C}>1$ population has a weak negative correlation with $R_{\text{HXR\_F}}$ and $R_{\text{HXR\_I*}}$; the $R_{C}<1$ population is not correlated with $R_{\text{HXR\_F}}$ nor $R_{\text{HXR\_I*}}$. The Mann-Whitney test indicates only modest or marginal population separation. Thus, the signed PVEC does not influence the electron precipitation as expected.

\subsection{Chi-square tests on the categorical association between photospheric parameters and HXR asymmetry}

We further test the null hypothesis that the stronger HXR footpoint is equally likely to be associated with a stronger (or weaker) photospheric parameter $V$. We categorized the corresponding ratio $R_V$ into two groups, i.e., $R_V >1$, where $V$ is larger at the stronger HXR footpoint FP1, and $R_V<1$, where $V$ is larger at the weaker footpoint FP2. The former (latter) group has $N_+$ ($N_-$) events, while the expected number of events $E_+=E_-=N/2$ for both groups under the null hypothesis of independence, with $N$ being the total number of events. So $\chi^2 = (N_+-E_+)^2 + (N_- - E_-)^2$, and the effect size $\mathrm{ES} = \sqrt{\chi^2/N}$.

Regarding magnetic parameters, we got $\chi^2\simeq 0.01$ for the ratio of magnetic field strength $R_{\text{UM\_B}}$, corresponding to a $p$-value of 0.92, and $\mathrm{ES\simeq0.01}$, which indicates a negligible association between $R_{\text{UM\_B}}$ and HXR footpoint asymmetry. This result is consistent with the weak correlation between $R_{\text{UM\_B}}$ and HXR asymmetry ratios (Figure~\ref{fig:HXR-UM}a,b), and with the Mann–Whitney U-test which suggests no significant difference between pro-MM ($R_{\text{UM\_B}} < 1$) and anti-MM ($R_{\text{UM\_B}} > 1$) cases. In contrast, for the unsigned magnetic flux ratio $R_{\text{UM\_F}}$, we got $\chi^2\simeq25.25$, corresponding to a $p$-value close to zero, which indicates that the anti-MM group ($R_{\text{UM\_F}}>1$) is aligned with the HXR footpoint asymmetry. The effect size of 0.50 suggests a moderate categorical association between the magnetic flux dominance and the HXR footpoint dominance. This result is consistent with the strong positive correlation between $R_{\text{UM\_F}}$ and HXR asymmetry ratios in the scatter plots (Figure~\ref{fig:HXR-UM}(c \& d)), but contrary to the classical magnetic mirroring expectation. 

Regarding electric current parameters, we have $\chi^2\simeq 0.01$ for the signed PVEC ratio $R_{\text{C}}$, corresponding to a $p$-value of 0.92, and $\mathrm{ES} \simeq 0.01$, which suggests a negligible categorical association between $R_{\text{C}}$ and HXR footpoint asymmetry. In contrast, the $\chi^2$ statistic for the unsigned PVEC ratio $R_{\text{UC}}$ is as large as 50, corresponding to a $p$-value close to zero, which strongly rejects the null hypothesis that $R_{\text{UC}}$ is independent of the HXR footpoint asymmetry. The effect size is as large as 0.70, indicating a robust categorical association dictating that the unsigned electric current is likely larger at the stronger HXR footpoint. The $\chi^2$ result is fully consistent with the strong Spearman correlation between $R_{\text{UC}}$ and HXR asymmetry ratios, as well as the Mann–Whitney U-test (Figure~\ref{fig:HXR-UC-EC}(c \& d)). Therefore, it is the electric-current magnitude rather than the electric-current polarity that is strongly linked to the HXR footpoint asymmetry. 

\begin{figure}
    \centering
    \includegraphics[width=0.3\linewidth]{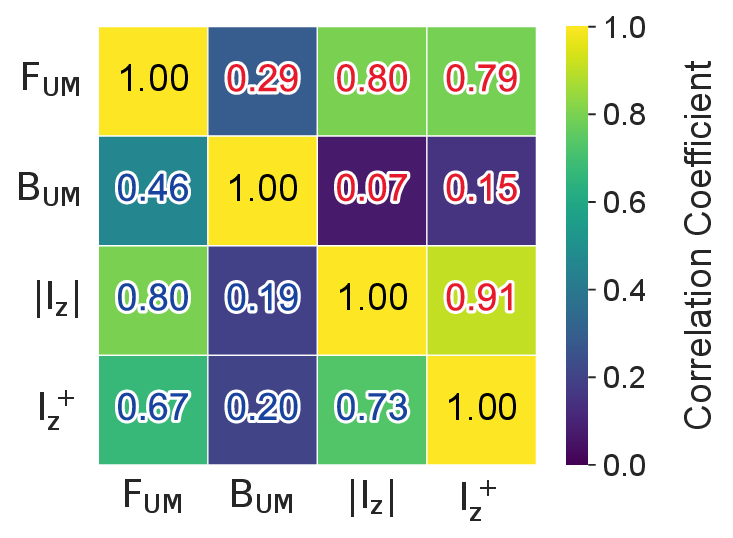}
    \caption{Correlation matrix between photospheric parameters at HXR footpoints. Pearson coefficient values are marked by red colors and Spearman's by blue colors. Cell's color represents the coefficient value accordingly to the color bar.} \label{fig:cc_FP_param}
\end{figure}

\section{Discussion} \label{sec:discussion} 

\subsection{HXR footpoint asymmetry in flux, brightness, and size} 
\label{subsec:discuss_asymmetry}

Figure~\ref{fig:HXRF-HXRI} demonstrates that the asymmetry between conjugate HXR footpoints is common and often large. In our sample, many events exhibit substantial brightness contrast between the conjugate footpoints; the distribution of the HXR flux ratio $R_{\text{HXR\_F}}$ and of the maximum brightness ratio $R_{\text{HXR\_I*}}$ do not cluster tightly around unity but instead are broadly spread and right-skewed, with a mean value of 4.03 and 1.39, respectively (Figure~\ref{fig:HXRF-HXRI}d). 

The relation displayed in Figure~\ref{fig:HXRF-HXRI}a between the HXR maximum brightness ratio $R_{\text{HXR\_I*}}$ and the flux ratio $R_{\text{HXR\_F}}$ can be well approximated by a linear function in the linear–logarithm coordinate system, i.e., $\log_\mathrm{10} R_\mathrm{HXR\_F} = a + b \, R_\text{HXR\_I*} $ with the best-fit coefficients $a\simeq -1.22$ and $b\simeq 1.20$ (black solid line in Figure~\ref{fig:HXRF-HXRI}a), which is equivalent to
\begin{equation*}
    \mathrm{R_\mathrm{HXR\_F}} 
    \simeq 
    0.06 \, \exp(2.76\,R_\text{HXR\_I*}),
\end{equation*}
which makes it explicit that modest changes in maximum brightness ratio correspond to exponential changes in integrated flux ratio, despite a potential underestimation of the recorded intensity due to the limited dynamic range of RHESSI. Further, we fit the data with a power-law function $R_{\mathrm{HXR\_F}} = a + b\,R_{\text{HXR\_I*}}^{\gamma}$ and an exponential function $R_{\mathrm{HXR\_F}} = a +  b\,\exp(c\,R_{\text{HXR\_I*}})$. The least-squares fitting results are given as follows: 
\begin{align} 
\begin{split}
R_{\mathrm{HXR\_F}} &\simeq 0.99 + 0.19\,R_{\text{HXR\_I*}}^{6.69}; \\
R_{\mathrm{HXR\_F}} &\simeq 0.64 + 0.01\exp(3.84\,R_{\text{HXR\_I*}}).\label{eq:flux_vs_intensity}
\end{split}
\end{align}
By excluding two outliers in the upper-right corner and including a constant $a$, the power-law ($\chi^2\simeq92.5$) and exponential ($\chi^2\simeq93.0$) fits significantly reduce the $\chi^2$ value as compared with the linear fit in linear-log space ($\chi^2\simeq266.6$). Thus, the nonlinear correlation between $R_{\mathrm{HXR\_F}}$ and $R_{\text{HXR\_I*}}$ can be equally well represented by power-law and exponential functions. Although the fitting coefficients are empirically robust for the chosen flare events and reconstruction method, it will not be surprising that they shift under alternative imaging algorithms or contour levels; for example, the footpoint regions as defined by different HXR contour levels give different mean magnetic field strengths \citep[][their Figure 1]{Goff2004}. Further, as $R_{\text{HXR\_F}}$ is linearly related to the source area ratio $R_{\text{HXR\_A}}$ (Figure~\ref{fig:HXRF-HXRI}b), the nonlinear coupling between $R_{\text{HXR\_F}}$ and $R_{\text{HXR\_I*}}$ is also translated to a similar correlation between $R_{\text{HXR\_A}}$ and $R_{\text{HXR\_I*}}$ (Figure~\ref{fig:HXRF-HXRI}c). 

The HXR footpoint asymmetry is also manifested as the asymmetry of source sizes. The HXR source size can be modulated by the mirroring effect and the pitch-angle distribution of the precipitating electrons \citep{Warmth&mann2013}. Electrons with a large pitch angle are mirrored higher while field aligned electrons penetrate deeper. Hence a more spread pitch-angle distribution makes the source larger. But for the same population of electrons precipitating towards the conjugate footpoints, they encounter stronger mirroring effect at the footpoint where magnetic field converges faster, which makes the HXR footpoint source both weaker and smaller. Thus, the mirroring effect may also contribute to the linear correlation between the footpoint flux ratio and the area ratio. 

Numerical simulations suggest that the HXR source size is mainly determined by the atmospheric density structure \citep{Battaglia2012}, but in observation the source morphology and size are also dependent on the instrument resolution and reconstruction algorithm \citep{Hurford2002, 2004SoPh..219..149A,Dennis2009,Warmth&mann2013}. Moreover, it is unclear whether the observed HXR sources are fully resolved. Often HXR footpoints appear compact but the associated UV/EUV flare ribbons are narrower and more elongated. Ribbon-like HXR footpoint sources are only occasionally reported \citep{Liu2007,Krucker2011}. This contrast suggests that the production of HXRs might be distributed over a number of distinct but transient sources along the flare ribbons. HXR imaging reconstruction typically integrates over seconds to tens of seconds, which naturally `blurs' out dynamic sources \citep{Qiu&Wang2006} and diminishes weak ones. The ratio-based approach adopted here may mitigate some but not all instrument/algorithm-driven spread.

Although previous studies almost unanimously define the HXR asymmetry in terms of flux ratios \citep[e.g.,][and references therein]{Yang2012}, it is important to emphasize that $R_{\text{HXR\_F}}$ and $R_{\text{HXR\_I*}}$ measure related (Figure~\ref{fig:HXRF-HXRI}a) but distinct physical quantities, and their differences carry diagnostic information.  $R_{\text{HXR\_F}}$ integrates emission over the whole reconstructed source area, therefore being sensitive to both the local intensity and the spatial extent of the precipitation region. $R_{\text{HXR\_I*}}$ captures the local intensity contrast, but is less sensitive to how extended or fragmented the emitting region is. Obviously, HXR flux and brightness are not interchangeable, but have different implications for source geometry and probably local precipitation efficiency. 

Further, due to the linear correlation between $R_{\text{HXR\_F}}$ and $R_{\text{HXR\_A}}$, $R_{\mathrm{HXR\_\langle I\rangle}}$ is rather balanced, i.e., the conjugate HXR footpoints are similar in mean brightness, which seems to suggest a plain scenario that the same population of accelerated electrons is simply split into two beams injecting into the flare loop, therefore accounting for the similar temporal and spectral characteristics \citep{Jin&Ding2007,Saint-Hilaire2008} as well as mean brightness, but different HXR production (flux), at the precipitation sites. However, it remains to be investigated whether the observed features are common among different reconstruction algorithms or unique to the CLEAN algorithm, because of the way `Clean components' are iteratively subtracted from the back projection (`dirty') map and then convoluted with the instrument Point Spread Function \cite[]{Hurford2002}. Other instrumental effects, such as the limited dynamic range of RHESSI imaging, may also need to be taken into account. 

\subsection{Physical mechanism of HXR footpoint asymmetry}
Instead of the anti-correlation between HXR asymmetry and photospheric magnetic field strength asymmetry at conjugate HXR footpoints as expected by the magnetic mirroring effect, we have demonstrated a significant positive correlation between HXR asymmetry and magnetic flux asymmetry, which excludes magnetic mirroring as an important mechanism in modulating the electron precipitation. We have also demonstrated a significant positive correlation between HXR asymmetry and unsigned electric-current asymmetry at conjugate HXR footpoints, which excludes the PVEC-associated potential drops as an important mechanism in modulating the electron precipitation. 

Between various photospheric parameter ratios and HXR asymmetry ratio, a consistent and clear hierarchy of association strength have been revealed as follows: $R_{\text{UC}} > R_{\text{UM\_F}} \gg  R_{\text{UM\_B}}$,  with $R_{\text{C}}$ showing negligible relevance, in terms of correlation (Pearson and Spearman), population separation (Mann-Whitney U-test), and categorical dependence ($\chi^2$ test). The stronger HXR footpoint is thus associated with larger unsigned magnetic flux and larger unsigned electric current through the footpoint region. However, magnetic flux and electric current are not independent of each other. In Figure~\ref{fig:cc_FP_param}, we present Spearman and Pearson correlation coefficients between the photospheric parameters for the individual HXR footpoints. The strongest correlations are found between three pairs: unsigned magnetic flux $F_{\text{UM}}$ and unsigned PVEC $|I_z|$ ($r_{\text{s}}$ and $r_{\text{p}}\approx0.80$), $|I_z|$ and $I_z^+$ ($r_{\text{s}}\approx0.73$ and $r_{\text{p}}\approx0.91$), $F_{\text{UM}}$ and $I_z^+$ ($r_{\text{s}}$ $\approx$ 0.67 and $r_{\text{p}}$ $\approx$ 0.79). The close agreement between Spearman and Pearson values indicates that these relations are not only monotonic but approximately linear over the sampled dynamic range. Thus, the unsigned PVEC may play a leading role in modulating the HXR asymmetry; other parameters are linked to the HXR asymmetry through their intimate relation with the unsigned PVEC. 

The strong positive correlation between HXR asymmetry and the asymmetry of unsigned PVEC at the conjugate footpoints suggests that PVECs not only trace magnetic shear and non-potentiality in active regions, but reflect the coronal reconnection environment. In the pre-eruption corona, where plasmas remain quasi-stationary and Lorentz forces dominate over plasma pressure and gravity, i.e., in a force-free condition $\mathbf{j}\times\mathbf{B}\approx 0$, the electric current circuits cannot close in the corona but are aligned along magnetic field, flowing into and out of the corona at the photosphere. In the 3D extension of the standard flare model \citep{Janvier2014,Janvier2017}, electric-current ribbons at the solar surface map the footprints of the coronal current layers formed at separatrices and QSLs, including the vertical current layer beneath the magnetic flux rope and the current layer wrapping around the flux rope.  The observation that accelerated electrons preferentially precipitate at the HXR footpoint with larger electric current suggests two possible scenarios: 1) flare-ribbon regions with higher PVECs are more likely connected to the electron acceleration site through electric current circuits; consequently more accelerated electrons are allocated for the stronger HXR footpoint. 2) such regions are more likely associated with the extremely dense currents required for the generation of anomalous resistivity by current-driven micro-turbulence \citep{Haerendel2017}, whose detection is far beyond the capabilities of modern instruments. Considering the resemblance in lightcurves, spectra, and mean brightness between conjugate HXR footpoints, we are leaning toward the 1st scenario, in which the HXR footpoint asymmetry is due to a redistribution of accelerated electrons into two different beams, rather than due to different acceleration processes.

We also noted in passing that HXR footpoint asymmetry could be both time and energy dependent \citep[e.g., Figure~\ref{fig:time-energy-depend}; see also][]{Yang2012}. It is worthy of further investigations on whether the temporal change of HXR asymmetry is related to the temporal change of photospheric parameters and whether the energy dependence of HXR asymmetry is related to the different photospheric parameters due to different source centroid/maximum locations.

\section{Conclusion} \label{sec:conclusion}

In this work, we performed a statistical investigation of HXR footpoint asymmetry in 67 M- and X-class double-ribbon flares. Our primary findings are:
\begin{itemize} 
     \item The conjugate HXR footpoints are asymmetric in source flux, maximum brightness, and sizes, but symmetric in mean brightness. The source flux ratio shows a strong linear correlation with the size ratio and a nonlinear correlation with the maximum brightness ratio.     
     \item The unsigned PVEC asymmetry exhibits a strong positive correlation with HXR footpoint asymmetry, indicating that electric-current-associated processes play an important role in modulating electron precipitation. However, the signed PVEC asymmetry is irrelevant to HXR footpoint asymmetry. 
     \item The magnetic field strength and flux asymmetry do not exhibit any anti-correlation but rather a positive correlation with HXR footpoint asymmetry, challenging the effectiveness of magnetic mirroring.    
\end{itemize}

Overall, our findings suggest that the asymmetry of conjugate HXR footpoints in double-ribbon flares is governed primarily by the electric current systems associated with magnetic reconnection, rather than by magnetic field convergence effects alone. In the framework of the 3D standard flare model, electric current asymmetry likely reflects asymmetric reconnection geometry, electric fields, or turbulence within QSLs and HFTs. Their interplay seems to control the spatial partitioning of nonthermal electrons in solar flares. 

\section*{Acknowledgments}

We would like to thank the RHESSI team for excellent data and software tools. This work was supported by the National Key R\&D Program of China (2022YFF0503002), the Strategic Priority Program of the Chinese Academy of Sciences (XDB0560102), and the NSFC (42274204, 42188101, 11925302). M.M.M. acknowledges support from the CAS--TWAS Fellowship Programme, jointly offered by the Chinese Academy of Sciences and The World Academy of Sciences. A.K.A. is supported by the National Science Centre, Poland, grant No. 2023/49/B/ST9/02409.

\appendix
\section{Mann-Whitney U-test for sample difference} \label{app:U_test}
We employed the Mann-Whitney U-test to evaluate whether significant differences exist between two samples with a different sizes, $n_1$ and $n_2$ \citep{nonparametric_book}. We rank the values from both samples together and then compare the sum of ranks $R_i$ to test if one sample has systematically higher (or lower) ranks than the other through the $U$ statistics:
\begin{equation}
    U_i=n_1 n_2+\frac{n_i (n_i +1)}{2}-\sum R_i, \quad i=1,\,2.
\end{equation}
The smaller $U_i$ is adopted to calculate the $z$-score under the normal approximation:
\begin{equation}
    z=\frac{U_i -\overline{x_U}}{S_U},
\end{equation}
with the mean $\overline{x_U}=n_1 n_2/2$ and the standard deviation $S_U=\sqrt{n_1 n_2 (n_1 + n_2)/12}$. At a significance level $\alpha = 0.05$, the null hypothesis that no significant difference exists between the two samples cannot be rejected if the $z$-score falls within the range of $-1.96 \leq z \leq 1.96$ for a two-tailed test. We further calculate the effect size (ES) to assess the degree of association between the two samples, i.e.,  
\begin{equation}
    \mathrm{ES}=\frac{|z|}{\sqrt{n}},
\end{equation}
where $n$ is the total number of observations. ES is categorized as small (0.1), medium (0.3), and large (0.5) by the conventions defined by \cite{Cohen}.

\newpage
\bibliography{references}{}

@INCOLLECTION{1964NASSP..50..451C,
       author = {{Carmichael}, H.},
        title = "{A Process for Flares}",
    booktitle = {NASA Special Publication},
         year = 1964,
       editor = {{Hess}, Wilmot N.},
       volume = {50},
        pages = {451},
       adsurl = {https://ui.adsabs.harvard.edu/abs/1964NASSP..50..451C}
}

@ARTICLE{1966Natur.211..695S,
       author = {{Sturrock}, P.~A.},
        title = "{Model of the High-Energy Phase of Solar Flares}",
      journal = {\nat},
         year = 1966,
        month = aug,
       volume = {211},
       number = {5050},
        pages = {695-697},
          doi = {10.1038/211695a0},
       adsurl = {https://ui.adsabs.harvard.edu/abs/1966Natur.211..695S}
}

@ARTICLE{1974SoPh...34..323H,
       author = {{Hirayama}, T.},
        title = "{Theoretical Model of Flares and Prominences. I: Evaporating Flare Model}",
      journal = {\solphys},
         year = 1974,
        month = feb,
       volume = {34},
       number = {2},
        pages = {323-338},
          doi = {10.1007/BF00153671},
       adsurl = {https://ui.adsabs.harvard.edu/abs/1974SoPh...34..323H}
}

@ARTICLE{1976SoPh...50...85K,
       author = {{Kopp}, R.~A. and {Pneuman}, G.~W.},
        title = "{Magnetic reconnection in the corona and the loop prominence phenomenon.}",
      journal = {\solphys},
         year = 1976,
        month = sep,
       volume = {50},
       number = {1},
        pages = {85-98},
          doi = {10.1007/BF00206193},
       adsurl = {https://ui.adsabs.harvard.edu/abs/1976SoPh...50...85K}
}

@ARTICLE{1971SoPh...18..489B,
       author = {{Brown}, John C.},
        title = "{The Deduction of Energy Spectra of Non-Thermal Electrons in Flares from the Observed Dynamic Spectra of Hard X-Ray Bursts}",
      journal = {\solphys},
         year = 1971,
        month = jul,
       volume = {18},
       number = {3},
        pages = {489-502},
          doi = {10.1007/BF00149070},
       adsurl = {https://ui.adsabs.harvard.edu/abs/1971SoPh...18..489B}
}

@ARTICLE{2003ApJ...595L..97H,
       author = {{Holman}, Gordon D. and {Sui}, Linhui and {Schwartz}, Richard A. and {Emslie}, A. Gordon},
        title = "{Electron Bremsstrahlung Hard X-Ray Spectra, Electron Distributions, and Energetics in the 2002 July 23 Solar Flare}",
      journal = {\apjl},
         year = 2003,
        month = oct,
       volume = {595},
       number = {2},
        pages = {L97-L101},
          doi = {10.1086/378488},
       adsurl = {https://ui.adsabs.harvard.edu/abs/2003ApJ...595L..97H}
}

@ARTICLE{2005A&A...435..743S,
       author = {{Saint-Hilaire}, P. and {Benz}, A.~O.},
        title = "{Thermal and non-thermal energies of solar flares}",
      journal = {\aap},
         year = 2005,
        month = may,
       volume = {435},
       number = {2},
        pages = {743-752},
          doi = {10.1051/0004-6361:20041918},
       adsurl = {https://ui.adsabs.harvard.edu/abs/2005A&A...435..743S}
}

@PHDTHESIS{Sakao1994PhDT,
       author = {{Sakao}, T.},
        title = "{Characteristics of solar flare hard X-ray sources as revealed with the Hard X-ray Telescope aboard the Yohkoh satellite}",
       school = {University of Tokyo, Japan},
         year = 1994,
        month = jun,
       adsurl = {https://ui.adsabs.harvard.edu/abs/1994PhDT.......335S}
}

@ARTICLE{Sakao1996,
       author = {{Sakao}, T. and {Kosugi}, T. and {Masuda}, S. and {Yaji}, K. and {Inda-Koide}, M. and {Makishima}, K.},
        title = "{Characteristics of hard X-ray double sources in impulsive solar flares}",
      journal = {Advances in Space Research},
         year = 1996,
        month = mar,
       volume = {17},
       number = {4-5},
        pages = {67-70},
          doi = {10.1016/0273-1177(95)00544-O},
       adsurl = {https://ui.adsabs.harvard.edu/abs/1996AdSpR..17d..67S}
}

@ARTICLE{Goff2004,
       author = {{Goff}, C.~P. and {Matthews}, S.~A. and {van Driel-Gesztelyi}, L. and {Harra}, L.~K.},
        title = "{Relating magnetic field strengths to hard X-ray emission in solar flares}",
      journal = {\aap},
         year = 2004,
        month = aug,
       volume = {423},
        pages = {363-372},
          doi = {10.1051/0004-6361:20040392},
       adsurl = {https://ui.adsabs.harvard.edu/abs/2004A&A...423..363G}
}

@ARTICLE{1984AdSpR...4b.153T,
       author = {{Trottet}, G. and {Vilmer}, N.},
        title = "{Electron spectra deduced from solar hard X-ray bursts}",
      journal = {Advances in Space Research},
         year = 1984,
        month = jan,
       volume = {4},
       number = {2-3},
        pages = {153-156},
          doi = {10.1016/0273-1177(84)90305-3},
       adsurl = {https://ui.adsabs.harvard.edu/abs/1984AdSpR...4b.153T}
}

@ARTICLE{Siarkowski&Falewicz2004,
       author = {{Siarkowski}, M. and {Falewicz}, R.},
        title = "{Variations of the hard X-ray footpoint asymmetry in a solar flare}",
      journal = {\aap},
         year = 2004,
        month = dec,
       volume = {428},
        pages = {219-226},
          doi = {10.1051/0004-6361:20041036},
       adsurl = {https://ui.adsabs.harvard.edu/abs/2004A&A...428..219S}
}

@ARTICLE{Falewicz&Siarkowski2007,
       author = {{Falewicz}, R. and {Siarkowski}, M.},
        title = "{On the causes of hard x-ray asymmetry in solar flares}",
      journal = {\aap},
         year = 2007,
        month = jan,
       volume = {461},
       number = {1},
        pages = {285-293},
          doi = {10.1051/0004-6361:20065670},
       adsurl = {https://ui.adsabs.harvard.edu/abs/2007A&A...461..285F}
}

@ARTICLE{Yang2012,
       author = {{Yang}, Ya-Hui and {Cheng}, C.~Z. and {Krucker}, S{\"a}m and {Hsieh}, Min-Shiu and {Chen}, Nai-Hwa},
        title = "{Asymmetry of Hard X-Ray Emissions at Conjugate Footpoints in Solar Flares}",
      journal = {\apj},
         year = 2012,
        month = sep,
       volume = {756},
       number = {1},
          eid = {42},
        pages = {42},
          doi = {10.1088/0004-637X/756/1/42},
       adsurl = {https://ui.adsabs.harvard.edu/abs/2012ApJ...756...42Y}
}

@ARTICLE{Aschwanden1999,
       author = {{Aschwanden}, Markus J. and {Fletcher}, Lyndsay and {Sakao}, Taro and {Kosugi}, Takeo and {Hudson}, Hugh},
        title = "{Deconvolution of Directly Precipitating and Trap-precipitating Electrons in Solar Flare Hard X-Rays. III.Yohkoh Hard X-Ray Telescope Data Analysis}",
      journal = {\apj},
         year = 1999,
        month = jun,
       volume = {517},
       number = {2},
        pages = {977-989},
          doi = {10.1086/307230},
       adsurl = {https://ui.adsabs.harvard.edu/abs/1999ApJ...517..977A}
}

@ARTICLE{2002SoPh..210....3L,
       author = {{Lin}, R.~P. and {Dennis}, B.~R. and {Hurford}, G.~J. and {Smith}, D.~M. and {Zehnder}, A. and {Harvey}, P.~R. and {Curtis}, D.~W. and {Pankow}, D. and {Turin}, P. and {Bester}, M. and {Csillaghy}, A. and {Lewis}, M. and {Madden}, N. and {van Beek}, H.~F. and {Appleby}, M. and {Raudorf}, T. and {McTiernan}, J. and {Ramaty}, R. and {Schmahl}, E. and {Schwartz}, R. and {Krucker}, S. and {Abiad}, R. and {Quinn}, T. and {Berg}, P. and {Hashii}, M. and {Sterling}, R. and {Jackson}, R. and {Pratt}, R. and {Campbell}, R.~D. and {Malone}, D. and {Landis}, D. and {Barrington-Leigh}, C.~P. and {Slassi-Sennou}, S. and {Cork}, C. and {Clark}, D. and {Amato}, D. and {Orwig}, L. and {Boyle}, R. and {Banks}, I.~S. and {Shirey}, K. and {Tolbert}, A.~K. and {Zarro}, D. and {Snow}, F. and {Thomsen}, K. and {Henneck}, R. and {Mchedlishvili}, A. and {Ming}, P. and {Fivian}, M. and {Jordan}, John and {Wanner}, Richard and {Crubb}, Jerry and {Preble}, J. and {Matranga}, M. and {Benz}, A. and {Hudson}, H. and {Canfield}, R.~C. and {Holman}, G.~D. and {Crannell}, C. and {Kosugi}, T. and {Emslie}, A.~G. and {Vilmer}, N. and {Brown}, J.~C. and {Johns-Krull}, C. and {Aschwanden}, M. and {Metcalf}, T. and {Conway}, A.},
        title = "{The Reuven Ramaty High-Energy Solar Spectroscopic Imager (RHESSI)}",
      journal = {\solphys},
         year = 2002,
        month = nov,
       volume = {210},
       number = {1},
        pages = {3-32},
          doi = {10.1023/A:1022428818870},
       adsurl = {https://ui.adsabs.harvard.edu/abs/2002SoPh..210....3L}
}

@ARTICLE{2004SoPh..219..149A,
       author = {{Aschwanden}, Markus J. and {Metcalf}, Thomas R. and {Krucker}, S{\"a}m and {Sato}, Jun and {Conway}, Andrew J. and {Hurford}, G.~J. and {Schmahl}, Edward J.},
        title = "{On the Photometric Accuracy of RHESSI Imaging and Spectrosocopy}",
      journal = {\solphys},
         year = 2004,
        month = jan,
       volume = {219},
       number = {1},
        pages = {149-157},
          doi = {10.1023/B:SOLA.0000021801.83038.aa},
archivePrefix = {arXiv},
       eprint = {astro-ph/0309499},
 primaryClass = {astro-ph},
       adsurl = {https://ui.adsabs.harvard.edu/abs/2004SoPh..219..149A}
}

@ARTICLE{Jin&Ding2007,
       author = {{Jin}, M. and {Ding}, M.~D.},
        title = "{Correlation and asymmetry between solar flare hard X-ray footpoints: a statistical study}",
      journal = {\aap},
         year = 2007,
        month = aug,
       volume = {471},
       number = {2},
        pages = {705-709},
          doi = {10.1051/0004-6361:20077202},
       adsurl = {https://ui.adsabs.harvard.edu/abs/2007A&A...471..705J}
}

@ARTICLE{Liu2009,
       author = {{Liu}, Wei and {Petrosian}, Vah{\'e} and {Dennis}, Brian R. and {Holman}, Gordon D.},
        title = "{Conjugate Hard X-Ray Footpoints in the 2003 October 29 X10 Flare: Unshearing Motions, Correlations, and Asymmetries}",
      journal = {\apj},
         year = 2009,
        month = mar,
       volume = {693},
       number = {1},
        pages = {847-867},
          doi = {10.1088/0004-637X/693/1/847},
archivePrefix = {arXiv},
       eprint = {0805.1055},
 primaryClass = {astro-ph},
       adsurl = {https://ui.adsabs.harvard.edu/abs/2009ApJ...693..847L}
}

@ARTICLE{2017ApJ...845...49K,
       author = {{Kazachenko}, Maria D. and {Lynch}, Benjamin J. and {Welsch}, Brian T. and {Sun}, Xudong},
        title = "{A Database of Flare Ribbon Properties from the Solar Dynamics Observatory. I. Reconnection Flux}",
      journal = {\apj},
         year = 2017,
        month = aug,
       volume = {845},
       number = {1},
          eid = {49},
        pages = {49},
          doi = {10.3847/1538-4357/aa7ed6},
archivePrefix = {arXiv},
       eprint = {1704.05097},
 primaryClass = {astro-ph.SR},
       adsurl = {https://ui.adsabs.harvard.edu/abs/2017ApJ...845...49K}
}

@ARTICLE{He2020,
       author = {{He}, Yuwei and {Liu}, Rui and {Liu}, Lijuan and {Chen}, Jun and {Wang}, Wensi and {Wang}, Yuming},
        title = "{Electric Currents through J-shaped and Non-J-shaped Flare Ribbons}",
      journal = {\apj},
         year = 2020,
        month = sep,
       volume = {900},
       number = {1},
          eid = {38},
        pages = {38},
          doi = {10.3847/1538-4357/aba52a},
archivePrefix = {arXiv},
       eprint = {2007.05693},
 primaryClass = {astro-ph.SR},
       adsurl = {https://ui.adsabs.harvard.edu/abs/2020ApJ...900...38H}
}

@ARTICLE{2006ApJ...651..553Z,
       author = {{Zharkova}, Valentina V. and {Gordovskyy}, Mykola},
        title = "{The Effect of the Electric Field Induced by Precipitating Electron Beams on Hard X-Ray Photon and Mean Electron Spectra}",
      journal = {\apj},
         year = 2006,
        month = nov,
       volume = {651},
       number = {1},
        pages = {553-565},
          doi = {10.1086/506423},
       adsurl = {https://ui.adsabs.harvard.edu/abs/2006ApJ...651..553Z}
}

@ARTICLE{Alexander&Metcalf2002,
       author = {{Alexander}, David and {Metcalf}, Thomas R.},
        title = "{Energy dependence of electron trapping in a solar flare}",
      journal = {\solphys},
         year = 2002,
        month = nov,
       volume = {210},
       number = {1},
        pages = {323-340},
          doi = {10.1023/A:1022457413628},
       adsurl = {https://ui.adsabs.harvard.edu/abs/2002SoPh..210..323A}
}

@ARTICLE{2012SoPh..275..207S,
       author = {{Scherrer}, P.~H. and {Schou}, J. and {Bush}, R.~I. and {Kosovichev}, A.~G. and {Bogart}, R.~S. and {Hoeksema}, J.~T. and {Liu}, Y. and {Duvall}, T.~L. and {Zhao}, J. and {Title}, A.~M. and {Schrijver}, C.~J. and {Tarbell}, T.~D. and {Tomczyk}, S.},
        title = "{The Helioseismic and Magnetic Imager (HMI) Investigation for the Solar Dynamics Observatory (SDO)}",
      journal = {\solphys},
         year = 2012,
        month = jan,
       volume = {275},
       number = {1-2},
        pages = {207-227},
          doi = {10.1007/s11207-011-9834-2},
       adsurl = {https://ui.adsabs.harvard.edu/abs/2012SoPh..275..207S}
}

@ARTICLE{2012SoPh..275..229S,
       author = {{Schou}, J. and {Scherrer}, P.~H. and {Bush}, R.~I. and {Wachter}, R. and {Couvidat}, S. and {Rabello-Soares}, M.~C. and {Bogart}, R.~S. and {Hoeksema}, J.~T. and {Liu}, Y. and {Duvall}, T.~L. and {Akin}, D.~J. and {Allard}, B.~A. and {Miles}, J.~W. and {Rairden}, R. and {Shine}, R.~A. and {Tarbell}, T.~D. and {Title}, A.~M. and {Wolfson}, C.~J. and {Elmore}, D.~F. and {Norton}, A.~A. and {Tomczyk}, S.},
        title = "{Design and Ground Calibration of the Helioseismic and Magnetic Imager (HMI) Instrument on the Solar Dynamics Observatory (SDO)}",
      journal = {\solphys},
         year = 2012,
        month = jan,
       volume = {275},
       number = {1-2},
        pages = {229-259},
          doi = {10.1007/s11207-011-9842-2},
       adsurl = {https://ui.adsabs.harvard.edu/abs/2012SoPh..275..229S}
}

@ARTICLE{2012SoPh..275...17L,
       author = {{Lemen}, James R. and {Title}, Alan M. and {Akin}, David J. and {Boerner}, Paul F. and {Chou}, Catherine and {Drake}, Jerry F. and {Duncan}, Dexter W. and {Edwards}, Christopher G. and {Friedlaender}, Frank M. and {Heyman}, Gary F. and {Hurlburt}, Neal E. and {Katz}, Noah L. and {Kushner}, Gary D. and {Levay}, Michael and {Lindgren}, Russell W. and {Mathur}, Dnyanesh P. and {McFeaters}, Edward L. and {Mitchell}, Sarah and {Rehse}, Roger A. and {Schrijver}, Carolus J. and {Springer}, Larry A. and {Stern}, Robert A. and {Tarbell}, Theodore D. and {Wuelser}, Jean-Pierre and {Wolfson}, C. Jacob and {Yanari}, Carl and {Bookbinder}, Jay A. and {Cheimets}, Peter N. and {Caldwell}, David and {Deluca}, Edward E. and {Gates}, Richard and {Golub}, Leon and {Park}, Sang and {Podgorski}, William A. and {Bush}, Rock I. and {Scherrer}, Philip H. and {Gummin}, Mark A. and {Smith}, Peter and {Auker}, Gary and {Jerram}, Paul and {Pool}, Peter and {Soufli}, Regina and {Windt}, David L. and {Beardsley}, Sarah and {Clapp}, Matthew and {Lang}, James and {Waltham}, Nicholas},
        title = "{The Atmospheric Imaging Assembly (AIA) on the Solar Dynamics Observatory (SDO)}",
      journal = {\solphys},
         year = 2012,
        month = jan,
       volume = {275},
       number = {1-2},
        pages = {17-40},
          doi = {10.1007/s11207-011-9776-8},
       adsurl = {https://ui.adsabs.harvard.edu/abs/2012SoPh..275...17L}
}

@ARTICLE{2012SoPh..275...41B,
       author = {{Boerner}, Paul and {Edwards}, Christopher and {Lemen}, James and {Rausch}, Adam and {Schrijver}, Carolus and {Shine}, Richard and {Shing}, Lawrence and {Stern}, Robert and {Tarbell}, Theodore and {Title}, Alan and {Wolfson}, C. Jacob and {Soufli}, Regina and {Spiller}, Eberhard and {Gullikson}, Eric and {McKenzie}, David and {Windt}, David and {Golub}, Leon and {Podgorski}, William and {Testa}, Paola and {Weber}, Mark},
        title = "{Initial Calibration of the Atmospheric Imaging Assembly (AIA) on the Solar Dynamics Observatory (SDO)}",
      journal = {\solphys},
         year = 2012,
        month = jan,
       volume = {275},
       number = {1-2},
        pages = {41-66},
          doi = {10.1007/s11207-011-9804-8},
       adsurl = {https://ui.adsabs.harvard.edu/abs/2012SoPh..275...41B}
}

@ARTICLE{Pesnell2012,
       author = {{Pesnell}, W. Dean and {Thompson}, B.~J. and {Chamberlin}, P.~C.},
        title = "{The Solar Dynamics Observatory (SDO)}",
      journal = {\solphys},
         year = 2012,
        month = jan,
       volume = {275},
       number = {1-2},
        pages = {3-15},
          doi = {10.1007/s11207-011-9841-3},
       adsurl = {https://ui.adsabs.harvard.edu/abs/2012SoPh..275....3P}
}

@ARTICLE{Saint-Hilaire2008,
       author = {{Saint-Hilaire}, P. and {Krucker}, S. and {Lin}, R.~P.},
        title = "{A Statistical Survey of Hard X-ray Spectral Characteristics of Solar Flares with Two Footpoints}",
      journal = {\solphys},
         year = 2008,
        month = jul,
       volume = {250},
       number = {1},
        pages = {53-73},
          doi = {10.1007/s11207-008-9193-9},
archivePrefix = {arXiv},
       eprint = {1111.4247},
 primaryClass = {astro-ph.SR},
       adsurl = {https://ui.adsabs.harvard.edu/abs/2008SoPh..250...53S}
}

@ARTICLE{2014SoPh..289.3483H,
       author = {{Hoeksema}, J. Todd and {Liu}, Yang and {Hayashi}, Keiji and {Sun}, Xudong and {Schou}, Jesper and {Couvidat}, Sebastien and {Norton}, Aimee and {Bobra}, Monica and {Centeno}, Rebecca and {Leka}, K.~D. and {Barnes}, Graham and {Turmon}, Michael},
        title = "{The Helioseismic and Magnetic Imager (HMI) Vector Magnetic Field Pipeline: Overview and Performance}",
      journal = {\solphys},
         year = 2014,
        month = sep,
       volume = {289},
       number = {9},
        pages = {3483-3530},
          doi = {10.1007/s11207-014-0516-8},
archivePrefix = {arXiv},
       eprint = {1404.1881},
 primaryClass = {astro-ph.SR},
       adsurl = {https://ui.adsabs.harvard.edu/abs/2014SoPh..289.3483H}
}

@ARTICLE{Haerendel2017,
       author = {{Haerendel}, G.},
        title = "{Evidence for Field-parallel Electron Acceleration in Solar Flares}",
      journal = {\apj},
         year = 2017,
        month = oct,
       volume = {847},
       number = {2},
          eid = {113},
        pages = {113},
          doi = {10.3847/1538-4357/aa8995},
       adsurl = {https://ui.adsabs.harvard.edu/abs/2017ApJ...847..113H}
}

@ARTICLE{Li1997,
       author = {{Li}, Jing and {Metcalf}, Thomas R. and {Canfield}, Richard C. and {W{\"u}lser}, Jean-Pierre and {Kosugi}, Takeo},
        title = "{What Is the Spatial Relationship between Hard X-Ray Footpoints and Vertical Electric Currents in Solar Flares?}",
      journal = {\apj},
         year = {1997},
        month = {jun},
       volume = {482},
       number = {1},
        pages = {490-497},
          doi = {10.1086/304131},
       adsurl = {https://ui.adsabs.harvard.edu/abs/1997ApJ...482..490L}
}

@ARTICLE{Canfield1993a,
       author = {{Canfield}, Richard C. and {de La Beaujardiere}, J.-F. and {Fan}, Yuhong and {Leka}, K.~D. and {McClymont}, A.~N. and {Metcalf}, Thomas R. and {Mickey}, Donald L. and {Wuelser}, Jean-Pierre and {Lites}, Bruce W.},
        title = "{The Morphology of Flare Phenomena, Magnetic Fields, and Electric Currents in Active Regions. I. Introduction and Methods}",
      journal = {\apj},
         year = 1993,
        month = jul,
       volume = {411},
        pages = {362},
          doi = {10.1086/172836},
       adsurl = {https://ui.adsabs.harvard.edu/abs/1993ApJ...411..362C}
}

@ARTICLE{Canfield1993b,
       author = {{Leka}, K.~D. and {Canfield}, Richard C. and {McClymont}, A.~N. and {de La Beaujardiere}, J.-F. and {Fan}, Yuhong and {Tang}, F.},
        title = "{The Morphology of Flare Phenomena, Magnetic Fields, and Electric Currents in Active Regions. II. NOAA Active Region 5747 (1989 October)}",
      journal = {\apj},
         year = 1993,
        month = jul,
       volume = {411},
        pages = {370},
          doi = {10.1086/172837},
       adsurl = {https://ui.adsabs.harvard.edu/abs/1993ApJ...411..370L}
}

@ARTICLE{Canfield1993c,
       author = {{de La Beaujardiere}, J.-F. and {Canfield}, Richard C. and {Leka}, K.~D.},
        title = "{The Morphology of Flare Phenomena, Magnetic Fields, and Electric Currents in Active Regions. III. NOAA Active Region 6233 (1990 August)}",
      journal = {\apj},
         year = 1993,
        month = jul,
       volume = {411},
        pages = {378},
          doi = {10.1086/172838},
       adsurl = {https://ui.adsabs.harvard.edu/abs/1993ApJ...411..378D}
}

@ARTICLE{Sharykin2020,
       author = {{Sharykin}, I.~N. and {Zimovets}, I.~V. and {Myshyakov}, I.~I.},
        title = "{Flare Energy Release at the Magnetic Field Polarity Inversion Line during the M1.2 Solar Flare of 2015 March 15. II. Investigation of Photospheric Electric Current and Magnetic Field Variations Using HMI 135 s Vector Magnetograms}",
      journal = {\apj},
         year = 2020,
        month = apr,
       volume = {893},
       number = {2},
          eid = {159},
        pages = {159},
          doi = {10.3847/1538-4357/ab84ef},
archivePrefix = {arXiv},
       eprint = {1905.03352},
 primaryClass = {astro-ph.SR},
       adsurl = {https://ui.adsabs.harvard.edu/abs/2020ApJ...893..159S}
}

@ARTICLE{Kundu1995,
       author = {{Kundu}, M.~R. and {Nitta}, N. and {White}, S.~M. and {Shibasaki}, K. and {Enome}, S. and {Sakao}, T. and {Kosugi}, T. and {Sakurai}, T.},
        title = "{Microwave and Hard X-Ray Observations of Footpoint Emission from Solar Flares}",
      journal = {\apj},
         year = 1995,
        month = nov,
       volume = {454},
        pages = {522},
          doi = {10.1086/176503},
       adsurl = {https://ui.adsabs.harvard.edu/abs/1995ApJ...454..522K}
}

@ARTICLE{Asai2002,
       author = {{Asai}, Ayumi and {Masuda}, Satoshi and {Yokoyama}, Takaaki and {Shimojo}, Masumi and {Isobe}, Hiroaki and {Kurokawa}, Hiroki and {Shibata}, Kazunari},
        title = "{Difference between Spatial Distributions of the H{\ensuremath{\alpha}} Kernels and Hard X-Ray Sources in a Solar Flare}",
      journal = {\apjl},
         year = 2002,
        month = oct,
       volume = {578},
       number = {1},
        pages = {L91-L94},
          doi = {10.1086/344566},
archivePrefix = {arXiv},
       eprint = {astro-ph/0209106},
 primaryClass = {astro-ph},
       adsurl = {https://ui.adsabs.harvard.edu/abs/2002ApJ...578L..91A}
}

@ARTICLE{Falewicz2006,
       author = {{Falewicz}, R. and {Siarkowski}, M. and {Berlicki}, A.},
        title = "{Hard X-ray emission at the footpoints of solar flares}",
      journal = {Advances in Space Research},
         year = 2006,
        month = jan,
       volume = {38},
       number = {5},
        pages = {956-961},
          doi = {10.1016/j.asr.2005.10.039},
       adsurl = {https://ui.adsabs.harvard.edu/abs/2006AdSpR..38..956F}
}

@ARTICLE{McClements&Alexander2005,
       author = {{McClements}, K.~G. and {Alexander}, D.},
        title = "{Fokker-Planck Modeling of Asymmetric Footpoint Hard X-Ray Emission in Solar Flares}",
      journal = {\apj},
         year = 2005,
        month = feb,
       volume = {619},
       number = {2},
        pages = {1153-1159},
          doi = {10.1086/426581},
       adsurl = {https://ui.adsabs.harvard.edu/abs/2005ApJ...619.1153M}
}

@ARTICLE{Daou&Alexander2016,
       author = {{Daou}, Antoun G. and {Alexander}, David},
        title = "{Hard X-Ray Asymmetry Limits in Solar Flare Conjugate Footpoints}",
      journal = {\apj},
         year = 2016,
        month = nov,
       volume = {832},
       number = {1},
          eid = {63},
        pages = {63},
          doi = {10.3847/0004-637X/832/1/63},
       adsurl = {https://ui.adsabs.harvard.edu/abs/2016ApJ...832...63D}
}

@ARTICLE{Holman2011,
       author = {{Holman}, G.~D. and {Aschwanden}, M.~J. and {Aurass}, H. and {Battaglia}, M. and {Grigis}, P.~C. and {Kontar}, E.~P. and {Liu}, W. and {Saint-Hilaire}, P. and {Zharkova}, V.~V.},
        title = "{Implications of X-ray Observations for Electron Acceleration and Propagation in Solar Flares}",
      journal = {\ssr},
         year = 2011,
        month = sep,
       volume = {159},
       number = {1-4},
        pages = {107-166},
          doi = {10.1007/s11214-010-9680-9},
archivePrefix = {arXiv},
       eprint = {1109.6496},
 primaryClass = {astro-ph.SR},
       adsurl = {https://ui.adsabs.harvard.edu/abs/2011SSRv..159..107H}
}

@ARTICLE{Fletcher2011,
       author = {{Fletcher}, L. and {Dennis}, B.~R. and {Hudson}, H.~S. and {Krucker}, S. and {Phillips}, K. and {Veronig}, A. and {Battaglia}, M. and {Bone}, L. and {Caspi}, A. and {Chen}, Q. and et al.},
        title = "{An Observational Overview of Solar Flares}",
      journal = {\ssr},
         year = 2011,
        month = sep,
       volume = {159},
       number = {1-4},
        pages = {19-106},
          doi = {10.1007/s11214-010-9701-8},
archivePrefix = {arXiv},
       eprint = {1109.5932},
 primaryClass = {astro-ph.SR},
       adsurl = {https://ui.adsabs.harvard.edu/abs/2011SSRv..159...19F}
}

@ARTICLE{Wang2017,
       author = {{Wang}, Wensi and {Liu}, Rui and {Wang}, Yuming and {Hu}, Qiang and {Shen}, Chenglong and {Jiang}, Chaowei and {Zhu}, Chunming},
        title = "{Buildup of a highly twisted magnetic flux rope during a solar eruption}",
      journal = {Nature Communications},
         year = 2017,
        month = nov,
       volume = {8},
          eid = {1330},
        pages = {1330},
          doi = {10.1038/s41467-017-01207-x},
       adsurl = {https://ui.adsabs.harvard.edu/abs/2017NatCo...8.1330W}
}

@ARTICLE{Gou2023,
       author = {{Gou}, Tingyu and {Liu}, Rui and {Veronig}, Astrid M. and {Zhuang}, Bin and {Li}, Ting and {Wang}, Wensi and {Xu}, Mengjiao and {Wang}, Yuming},
        title = "{Complete replacement of magnetic flux in a flux rope during a coronal mass ejection}",
      journal = {Nature Astronomy},
         year = 2023,
        month = jul,
       volume = {7},
        pages = {815-824},
          doi = {10.1038/s41550-023-01966-2},
archivePrefix = {arXiv},
       eprint = {2305.03217},
 primaryClass = {astro-ph.SR},
       adsurl = {https://ui.adsabs.harvard.edu/abs/2023NatAs...7..815G}
}

@ARTICLE{Janvier2014,
       author = {{Janvier}, M. and {Aulanier}, G. and {Bommier}, V. and {Schmieder}, B. and {D{\'e}moulin}, P. and {Pariat}, E.},
        title = "{Electric Currents in Flare Ribbons: Observations and Three-dimensional Standard Model}",
      journal = {\apj},
         year = 2014,
        month = jun,
       volume = {788},
       number = {1},
          eid = {60},
        pages = {60},
          doi = {10.1088/0004-637X/788/1/60},
archivePrefix = {arXiv},
       eprint = {1402.2010},
 primaryClass = {astro-ph.SR},
       adsurl = {https://ui.adsabs.harvard.edu/abs/2014ApJ...788...60J}
}

@ARTICLE{Musset2015,
       author = {{Musset}, S. and {Vilmer}, N. and {Bommier}, V.},
        title = "{Hard X-ray emitting energetic electrons and photospheric electric currents}",
      journal = {\aap},
         year = 2015,
        month = aug,
       volume = {580},
          eid = {A106},
        pages = {A106},
          doi = {10.1051/0004-6361/201424378},
archivePrefix = {arXiv},
       eprint = {1506.02724},
 primaryClass = {astro-ph.SR},
       adsurl = {https://ui.adsabs.harvard.edu/abs/2015A&A...580A.106M}
}

@ARTICLE{Janvier2017,
       author = {{Janvier}, Miho},
        title = "{Three-dimensional magnetic reconnection and its application to solar flares}",
      journal = {Journal of Plasma Physics},
         year = 2017,
        month = feb,
       volume = {83},
       number = {1},
          eid = {535830101},
        pages = {535830101},
          doi = {10.1017/S0022377817000034},
archivePrefix = {arXiv},
       eprint = {1612.06513},
 primaryClass = {astro-ph.SR},
       adsurl = {https://ui.adsabs.harvard.edu/abs/2017JPlPh..83a5301J}
}

@ARTICLE{Aschwanden2002,
       author = {{Aschwanden}, Markus J.},
        title = "{Particle acceleration and kinematics in solar flares - A Synthesis of Recent Observations and Theoretical Concepts (Invited Review)}",
      journal = {\ssr},
         year = 2002,
        month = jan,
       volume = {101},
       number = {1},
        pages = {1-227},
          doi = {10.1023/A:1019712124366},
       adsurl = {https://ui.adsabs.harvard.edu/abs/2002SSRv..101....1A}
}

@ARTICLE{Liu2007,
       author = {{Liu}, Chang and {Lee}, Jeongwoo and {Gary}, Dale E. and {Wang}, Haimin},
        title = "{The Ribbon-like Hard X-Ray Emission in a Sigmoidal Solar Active Region}",
      journal = {\apjl},
         year = 2007,
        month = apr,
       volume = {658},
       number = {2},
        pages = {L127-L130},
          doi = {10.1086/513739},
archivePrefix = {arXiv},
       eprint = {astro-ph/0702326},
 primaryClass = {astro-ph},
       adsurl = {https://ui.adsabs.harvard.edu/abs/2007ApJ...658L.127L}
}

@ARTICLE{Qiu&Wang2006,
       author = {{Qiu}, Jiong and {Wang}, Haimin},
        title = "{On the Temporal and Spatial Properties of Elementary Bursts}",
      journal = {\solphys},
         year = 2006,
        month = jul,
       volume = {236},
       number = {2},
        pages = {293-311},
          doi = {10.1007/s11207-006-0050-4},
       adsurl = {https://ui.adsabs.harvard.edu/abs/2006SoPh..236..293Q}
}

@ARTICLE{Krucker2011,
       author = {{Krucker}, S{\"a}m and {Hudson}, H.~S. and {Jeffrey}, N.~L.~S. and {Battaglia}, M. and {Kontar}, E.~P. and {Benz}, A.~O. and {Csillaghy}, A. and {Lin}, R.~P.},
        title = "{High-resolution Imaging of Solar Flare Ribbons and Its Implication on the Thick-target Beam Model}",
      journal = {\apj},
         year = 2011,
        month = oct,
       volume = {739},
       number = {2},
          eid = {96},
        pages = {96},
          doi = {10.1088/0004-637X/739/2/96},
       adsurl = {https://ui.adsabs.harvard.edu/abs/2011ApJ...739...96K}
}

@ARTICLE{Warmth&mann2013,
       author = {{Warmuth}, A. and {Mann}, G.},
        title = "{Thermal and nonthermal hard X-ray source sizes in solar flares obtained from RHESSI observations. I. Observations and evaluation of methods}",
      journal = {\aap},
         year = 2013,
        month = apr,
       volume = {552},
          eid = {A86},
        pages = {A86},
          doi = {10.1051/0004-6361/201219354},
       adsurl = {https://ui.adsabs.harvard.edu/abs/2013A&A...552A..86W}
}

@ARTICLE{Dennis2009,
       author = {{Dennis}, Brian R. and {Pernak}, Rick L.},
        title = "{Hard X-Ray Flare Source Sizes Measured with the Ramaty High Energy Solar Spectroscopic Imager}",
      journal = {\apj},
         year = 2009,
        month = jun,
       volume = {698},
       number = {2},
        pages = {2131-2143},
          doi = {10.1088/0004-637X/698/2/2131},
       adsurl = {https://ui.adsabs.harvard.edu/abs/2009ApJ...698.2131D}
}

@ARTICLE{Battaglia2012,
       author = {{Battaglia}, M. and {Kontar}, E.~P. and {Fletcher}, L. and {MacKinnon}, A.~L.},
        title = "{Numerical Simulations of Chromospheric Hard X-Ray Source Sizes in Solar Flares}",
      journal = {\apj},
         year = 2012,
        month = jun,
       volume = {752},
       number = {1},
          eid = {4},
        pages = {4},
          doi = {10.1088/0004-637X/752/1/4},
archivePrefix = {arXiv},
       eprint = {1204.1151},
 primaryClass = {astro-ph.SR},
       adsurl = {https://ui.adsabs.harvard.edu/abs/2012ApJ...752....4B}
}

@ARTICLE{Cohen,
       author = {{Cohen}, J.},
        title = "{Statistical Power Analysis for the Behavioral Sciences (2nd ed.).}",
      journal = {Routledge},
         year = 1988,
          doi = {10.4324/9780203771587},
}

@Book{nonparametric_book,
  title     = {Nonparametric statistics for non-statisticians},
  publisher = {John Wiley \& Sons, Inc.},
  year      = {2011},
  author    = {{Corder}, Gregory W and {Foreman}, Dale I},
}

@ARTICLE{Sharykin&Kosovichev2014,
       author = {{Sharykin}, I.~N. and {Kosovichev}, A.~G.},
        title = "{Fine Structure of Flare Ribbons and Evolution of Electric Currents}",
      journal = {\apjl},
         year = 2014,
        month = jun,
       volume = {788},
       number = {1},
          eid = {L18},
        pages = {L18},
          doi = {10.1088/2041-8205/788/1/L18},
archivePrefix = {arXiv},
       eprint = {1404.5104},
 primaryClass = {astro-ph.SR},
       adsurl = {https://ui.adsabs.harvard.edu/abs/2014ApJ...788L..18S}
}

@article{Sharykin_2015,
doi = {10.1088/0004-637X/808/1/72},
url = {https://doi.org/10.1088/0004-637X/808/1/72},
year = {2015},
month = {jul},
publisher = {The American Astronomical Society},
volume = {808},
number = {1},
pages = {72},
author = {Sharykin, I. N. and Kosovichev, A. G.},
title = {DYNAMICS OF ELECTRIC CURRENTS, MAGNETIC FIELD TOPOLOGY, AND HELIOSEISMIC RESPONSE OF A SOLAR FLARE},
journal = {The Astrophysical Journal}
}

@INPROCEEDINGS{Zimovets&Sharykin2022,
       author = {{Zimovets}, I. and {Sharykin}, I.},
        title = "{A brief review on vertical electric currents in flaring active regions at the Sun}",
    booktitle = {Astronomy at the Epoch of Multimessenger Studies},
         year = 2022,
        month = jan,
        pages = {42-46},
          doi = {10.51194/VAK2021.2022.1.1.006},
       adsurl = {https://ui.adsabs.harvard.edu/abs/2022aems.conf...42Z}
}

@ARTICLE{Zimovets2020apj,
       author = {{Zimovets}, I.~V. and {Sharykin}, I.~N. and {Gan}, W.~Q.},
        title = "{Relationships between Photospheric Vertical Electric Currents and Hard X-Ray Sources in Solar Flares: Statistical Study}",
      journal = {\apj},
         year = 2020,
        month = mar,
       volume = {891},
       number = {2},
          eid = {138},
        pages = {138},
          doi = {10.3847/1538-4357/ab75be},
archivePrefix = {arXiv},
       eprint = {2002.06646},
 primaryClass = {astro-ph.SR},
       adsurl = {https://ui.adsabs.harvard.edu/abs/2020ApJ...891..138Z}
}

@ARTICLE{Qiu&Gary2003,
       author = {{Qiu}, Jiong and {Gary}, Dale E.},
        title = "{Flare-related Magnetic Anomaly with a Sign Reversal}",
      journal = {\apj},
         year = 2003,
        month = dec,
       volume = {599},
       number = {1},
        pages = {615-625},
          doi = {10.1086/379146},
       adsurl = {https://ui.adsabs.harvard.edu/abs/2003ApJ...599..615Q}
}

@ARTICLE{Burtseva2015,
       author = {{Burtseva}, O. and {Mart{\'\i}nez-Oliveros}, J.~C. and {Petrie}, G.~J.~D. and {Pevtsov}, A.~A.},
        title = "{Hard X-Ray Emission During Flares and Photospheric Field Changes}",
      journal = {\apj},
         year = 2015,
        month = jun,
       volume = {806},
       number = {2},
          eid = {173},
        pages = {173},
          doi = {10.1088/0004-637X/806/2/173},
archivePrefix = {arXiv},
       eprint = {1505.00509},
 primaryClass = {astro-ph.SR},
       adsurl = {https://ui.adsabs.harvard.edu/abs/2015ApJ...806..173B}
}

@ARTICLE{Hurford2002,
       author = {{Hurford}, G.~J. and {Schmahl}, E.~J. and {Schwartz}, R.~A. and {Conway}, A.~J. and {Aschwanden}, M.~J. and {Csillaghy}, A. and {Dennis}, B.~R. and {Johns-Krull}, C. and {Krucker}, S. and {Lin}, R.~P. and et al.},
        title = "{The RHESSI Imaging Concept}",
      journal = {\solphys},
         year = 2002,
        month = nov,
       volume = {210},
       number = {1},
        pages = {61-86},
          doi = {10.1023/A:1022436213688},
       adsurl = {https://ui.adsabs.harvard.edu/abs/2002SoPh..210...61H}
}
\bibliographystyle{aasjournal}

\end{document}